\documentclass[sn-basic]{sn-jnl}

\jyear{2022}%

\theoremstyle{thmstyleone}%
\theoremstyle{thmstyletwo}%

\theoremstyle{thmstylethree}%
\usepackage{caption}
\makeatletter
\newcommand\specialcaption[1]{%
  \@namedef{the\@captype}{#1}%
  \addtocounter{\@captype}{-1}\caption}
\makeatother

\begin{document}

\title[Scale Free Avalanches in Excitatory-Inhibitory Populations of Spiking Neurons]{Scale Free Avalanches in Excitatory-Inhibitory Populations of Spiking Neurons with Conductance Based Synaptic Currents}


\author*[1]{\fnm{Masud} \sur{Ehsani}}\email{masud.ehsani@mis.mpg.de}

\author[1,2]{\fnm{J\"{u}rgen} \sur{ Jost}}\email{jjost@mis.mpg.de}

\affil*[1]{ \orgname{Max Planck Institute for Mathematics in Sciences}, \orgaddress{\street{Inselstr.22}, \city{Leipzig}, \postcode{04103}, \state{Saxony}, \country{Germany }}}

\affil[2]{\orgname{Santa Fe Institute}, \orgaddress{\street{1399 Hyde Park Rd}, \city{Santa Fe}, \postcode{NM 87501}, \country{United States }}}


\abstract{We investigate spontaneous critical dynamics of excitatory and inhibitory (EI)
sparsely connected populations of spiking leaky integrate-and-fire neurons with 
conductance-based synapses. We use a bottom-up approach to derive a single
neuron gain function and a linear Poisson neuron approximation which we use to
study mean-field dynamics of the EI population and its bifurcations. In the low firing rate regime, the quiescent state
loses stability due to  saddle-node or Hopf bifurcations. In particular, at
the Bogdanov-Takens (BT) bifurcation point which is the intersection of the
Hopf bifurcation and the saddle-node bifurcation lines of the 2D dynamical system, the network shows avalanche dynamics with power-law avalanche size and duration distributions. This matches the characteristics of low firing spontaneous activity in the cortex. By linearizing gain functions and excitatory and inhibitory nullclines, we can approximate the location of the BT bifurcation point. This point in the control parameter phase space corresponds to the internal balance of excitation and inhibition and a slight excess of external excitatory input to the excitatory population. Due to the tight balance of average excitation and inhibition currents, the firing of the individual cells is fluctuation-driven. Around the BT point, the spiking of neurons is a Poisson process and the population average membrane potential of neurons is approximately at the middle of the operating interval $[V_{Rest}, V_{th}]$. Moreover, the EI network is close to both oscillatory and active-inactive phase transition regimes.}


\maketitle

\section{Introduction}\label{sec1}

Experiments have shown that in the absence of stimuli, the cortical population of neurons shows rich dynamical patterns, called spontaneous activity, which do not look random and entirely noise-driven but are structured in spatiotemporal patterns (\cite{Takeda16,Thompson14}). Spontaneous activity is assumed to be the substrate or background state of the neural system with functional significance (\cite{Raichle10}). Experimental findings on different temporal and spatial resolutions highlight the scale-free characteristic of spontaneous activity.
	
In microcircuits of the brain during spontaneous activity, we observe
avalanche dynamics. This mode of activity was first closely investigated by
\cite{Beggs03} in cultured slices of rat cortex using a
multi-electrode array with an inter-electrode distance of $200 \mu m$ to
record local field potentials (LFP).  An avalanche is defined as almost
synchronized epochs of activity separated by usually long periods of
inactivity. At higher temporal resolution this seemingly synchronized pattern
appears as a cascade of activity in micro-electrodes arrays initiated from one
(or a few) local sites that propagate through the network and finally terminate. 
The main finding of this seminal experimental paper is power-law scaling of
the probability density function for size and duration of
avalanches. And causing an excitation and
inhibition imbalance by injecting specific drugs destroys power-law scaling.  Further
studies confirm these results in different setups like  awake monkeys
(\cite{Petermann09}), in the cerebral cortex and hippocampus
of anesthetized, asleep, and awake rats ( \cite{Ribeiro10})
and the visual cortex of an anesthetized cat (\cite{Hahn10}). Besides LFP data several studies report the scale-free
avalanche size distribution based on spike data ( \cite{Friedman12},\cite{Hahn10}, \cite{Mazzoni07}).  \cite{Friedman12} analyzed
cultured slices of cortical tissue and collected data at individual neurons
with different spacing.  \cite{Klaus11} showed that a power law is the best fit for neural avalanches collected from in vivo and in vitro experiments.

Besides power-law scaling of size and duration of avalanches with exponents
$\tau \sim -1.5$ and $\alpha \sim -2$, respectively, they showed that the
temporal profile of avalanches is described by a single universal scaling
function. Average size versus average duration of avalanches is also a
power-law with $\langle s\rangle  = \langle T\rangle^{\dfrac{1}{\sigma \nu z}}$ linked by a scaling
relation    $\dfrac{\alpha-1}{\tau-1} = \dfrac{1}{\sigma \nu z}$ between
exponents.  In addition, the mean temporal profile of avalanches follows a
scaling form as in non-equilibrium critical dynamics, 
\begin{align} 
 S(t,T) \sim T^{1/\sigma \nu z -1} F(t/T)
	\end{align}	
Data sets collapse to the scaling function very well. 
 The appearance of power laws, scaling relations among their exponents, data
 collapse, and sensitivity to the imbalance of excitation and inhibition led
 to the hypothesis that somehow the brain is poised near criticality by a
 self-organization mechanism with the balance of excitatory and inhibitory
 rates as the self-organizing parameter. In this direction, many models have
 been presented in the past decades. Short-term plasticity in excitatory
 neuronal models has been investigated as a self-organizing principle for a
 non-conservative neuronal model \cite{Levina09, Levina07, Peng13,
   Santo17,Brochini16}. In addition, self-organization by other control
 parameters like degree of connectivity or synaptic strength \cite{Bornholdt03,Rybarsch14}, STDP \cite{Meisel09}, and balanced input \cite{Benayoun10} has been studied.

On the other hand, the
spontaneous firing of single neocortical neurons is
 considered to be a noisy, stochastic
process resembling a Poisson point process. It has been claimed that the balance of excitation and inhibition is a necessary condition for the noisy irregular firing of individual neurons as well as scale free avalanche patterns at the population level. Since the network is settled in a balanced state, a small deviation in the
balance condition leads to a local change in the firing rate. Therefore, the system is highly sensitive to input while maintaining a low firing rate and highly variable spike trains at the individual neuron level.  
 Inhibitory-excitatory balance can lead to  asynchronous cortical states in
 local populations  (\cite{Brunel99,Brunel00,Brunel08})  and the emergence of
 waves and fronts at a larger scale of cortical activity
 (\cite{Ermentrout98,Bressloff11}). At the level of individual neurons, this
 balance leads to highly irregular firing of neurons with inter-spike interval
 distribution with CV close to one and thus resembling a Poisson process (\cite{Softky93}). Studies using the voltage clamp method tracking conductance of excitatory and inhibitory synapses on neurons both in vivo and in vitro, confirmed that there exist proportionality and balance of inhibitory and excitatory currents during upstate (\cite{Haider06}), sensory input (\cite{Shu03}) and spontaneous activity (\cite{Okun08}). \cite{Benayoun10} proposed a stochastic model of spiking neurons which matches the Wilson-Cowan mean field in the limit of infinite system size that shows scale-free avalanches in the balanced state in which sum of excitation and inhibition is much larger than the net difference between them. Under symmetry condition on weights this makes the Jacobian to have two negative eigenvalues close to zero in the balanced state suggesting the system operating in the vicinity of a Bogdanov-Takens bifurcation point. \cite{Cowan13} used the method of path integral representation in the stochastic model of spiking neurons supplemented by anti-Hebbian synaptic plasticity as the self-organizing mechanism. Their network possesses bistability close to the saddle-node bifurcation point which is the origin of the avalanche behavior in the system.

In this work, we start from a bottom-up approach by analytically investigating
conditions on Poisson firing at the single neuron level and introducing
conditions on the balance of inhibitory and excitatory currents.  Next, we
build a linear Poisson neuron model with minimal error in the low firing rate
regime. The linear Poisson regime of firing is a segment of the dynamical
regime of the neuron response. We can use this linearization to form an
approximate linearized gain function. This gain function can then be used to
investigate dynamics of a sparsely or all to all connected homogeneous network
of inhibitory and excitatory neurons and its bifurcation diagram. We also
introduce another compatible approximation of gain functions by sigmoids. We
observe avalanche patterns with power-law distributed sizes and duration at
the intersection of saddle-node and Hopf bifurcation lines, i.e.,  at a
Bogdanov-Takens  (BT) bifurcation point of the mean field equations. At this point, the balance of
excitatory and inhibitory inputs leads to stationary values of membrane
potentials that allow Poisson firing at the single neuron level and avalanche
type dynamics at the population level. 
The firing of neurons is due to the accumulation of internal currents, and
external input by itself does not suffice to trigger firing. However, external
input imbalance to excitatory and inhibitory populations is needed for the
initiation of the avalanche. During each avalanche at the BT point, each
neuron on average activates one another neuron which leads to termination of
avalanches with power-law distributed durations and sizes. This is the case
when currents to single cells are balanced in a way that excess excitation
firing is compensated by inhibitory feedback. A linear relation between
excitatory and inhibitory rates close to the BT point enables us to write down
the dynamics of the excitatory population as a branching process. Close to the
BT point the branching parameter is close to one which is indicative of the
critical state.  Tuning the system at BT can be
attained  by a balance of inhibitory feedback leading to a condition on
synaptic weights and adjustment of excess external drive to the excitatory
population. This is investigated in another article ( \cite{Ehsani2}), where we show how learning by STDP and homeostatic synaptic plasticity as self-organizing principles can tune the system close to the BT point by regulating the inhibitory feedback strength and excitatory population gain.

\section{Neuron model and network architechture}

We use an integrate and fire neuron model in which the change in the membrane voltage of the neuron receiving time dependent synaptic current $i(t)$  follows : 
  \begin{align}
	C \dfrac{d v(t)}{dt} = g_{Leak}( v_{Leak} - v(t)) + i(t),
	\end{align}
for $v(t)<v_{th}$ . When the membrane voltage reaches $v_{th} = -50mv$, the  neuron spikes and immediately its membrane voltage resets to $v_{rest}$ which is equal to $v_{Leak} = -65mv$.

In the following, we want to concentrate on a model with just one type of inhibitory and one type of excitatory synapses, which can be seen as the average effect of the two types of synapses. We can write the synaptic inhibitory and excitatory current as 	
	\begin{align}
	 i(t) =  g_{inh}(t) * ( V_{Rinh} - v(t)) + g_{exc}(t) * (V_{Rexc} - v(t))
	\end{align}	
$V_{Rinh}$ and $V_{Rexc}$ are the reverse potentials of excitatory and inhibitory ion channels, and based on experimental studies we choose values of $-80mv$ and $0mv$ for them respectively. 
 $g_{inh}(t)$ and  $g_{exc}(t)$ are the conductances of inhibitory and
 excitatory ion channels. These conductances are changing by the inhibitory
 and excitatory input to the cell. Each spike of a presynaptic inhibitory or
 excitatory neuron $j$ to a postsynaptic neuron $k$ that is received by  $k$
 at time $t_0$  will change the inhibitory or excitatory ion channel 
 conductance of the postsynaptic neuron for $t>t_0$ according to  
 \begin{align}
	& g_{Inh}^k(t) = w_{kj} * g_{0}^{inh} * exp(-\dfrac{t-t_0}{\tau_{syn}^{inh}}) \nonumber \\
	&g_{Exc}^k(t) = w_{kj} * g_{0}^{exc} * exp(-\dfrac{t-t_0}{\tau_{syn}^{exc}})
	\end{align}	
 Here we assume that the rise time of synaptic conductances is very small compared to other time scales in the model and therefore, we modeled the synaptic current by a decay term with synaptic decay time constant $\tau_{syn}$ which we assume to be the same value of $5ms$ for both inhibitory and excitatory synapses.
In the remainder of this work, in the simulation, we consider a population of
$N_{Exc} =2*10^4$ and $N_{Inh} = 0.25*N_{Exc}$ inhibitory spiking neurons with
conductance-based currents introduced in this section. Each excitatory neuron
in the population is randomly connected to $k_{EE} = \dfrac{N_{Exc}}{100}=200$
excitatory and $k_{EI} = \dfrac{k_{EE}}{4}$ inhibitory neurons and each
inhibitory neuron is connected to $k_{IE} = k_{II} = \dfrac{k_{EE}}{4} $
excitatory and inhibitory neurons. The weights of excitatory synaptic
connections are in a range that $10-20$ synchronous excitatory spikes suffice
to depolarize the target neuron to the level of its firing threshold when it is initially at rest at the time of input arrival. Weights are being drawn from a log-normal probability density with low variance. Therefore, approximately $O(\sqrt{k_{EE}})$ spikes are adequate for firing. Assuming homogeneity in the population as we have discussed in the introduction we can build a mean-field equation for the excitatory and inhibitory population in this sparse network, assuming each neuron receives input with the same statistics.

\section{Results}\label{sec3}

\subsection{Response of a single neuron to the Poisson input} \label{sec3-sub1}

In this section, we want to consider the response of the neuron to a specific type of current, namely Poisson input. The reason to consider this type of input is that in an asynchronous firing state neurons receive Poisson input from other neurons.
  Assume that the number of afferents to each neuron is high  and the population
  activity is nearly constant with firing rate $r$. Assuming homogeneity in
  the number of connections and weights, then at any moment the probability
  distribution that for a neuron,  $k$ presynaptic  neurons out of a
  total number of $n$ presynaptic neurons are active is a binomial $f(n,k,R)= \begin{bmatrix}
           n\\
           k \\
         \end{bmatrix} R^k(1-R)^{n-k}$ which in the regime  $R<<1$ is well
         approximated by a Poisson distribution with parameter $nR$.

We first study the response of the neuron to a non-fluctuating constant
periodic synaptic current. Suppose the target neuron receives  constant
numbers $\rho _E$ and $\rho _I$ of excitatory and inhibitory spikes per unit
time,   with all the excitatory spikes having the same strength $w_E$ and all
the inhibitory spikes having the strength $w_I$. The conductance of the excitatory channels $g_{exc}(t) $ is modified by excitatory spikes arriving at times $ s<t$  :
\begin{align}
	 g_{exc}(t)  =  \int_{-\infty}^{t} g_{exc}^0 w_E \rho_E exp(-\dfrac{t-s}{\tau_{syn}^{exc}}) ds  =   g_{exc}^0 w_E \rho_E \tau_{syn}^{exc}
	\end{align}	
The same formula applies for the constant inhibitory  current. The potential of the target neuron fed by this current will reach a stationary value. If this stationary limit is greater than $V_{th}$ then the target neuron will fire periodically. This constraint reads as : 
 
\begin{align}
\rho_I <  \dfrac{g_{leak}*(V_{th} - V_{rest}) + g_{exc}^0*w_E*\rho _E*\tau * V_{th}} {g_{inh}^0*w_I*\tau(V_{inh} - V_{th}) } \label{eq:Cond}
\end{align}	

The stationary limit of the potential is a weighted average of reverse potentials,

\begin{align}
V_{st} = \dfrac{g_{L}V_{L} + g_{exc}^0 w_E \rho_E \tau V_{Rexc} + g_{inh}^0 w_I \rho_I \tau V_{Rinh}  }{g_{L} + g_{exc}^0 w_E \rho_E \tau+ g_{inh}^0 w_I \rho_I \tau} \label{eq:VST}
\end{align}	

\begin{figure}
				\centering
			\includegraphics[width=1\linewidth,trim={0cm, 0cm, 0cm, 0cm},clip]{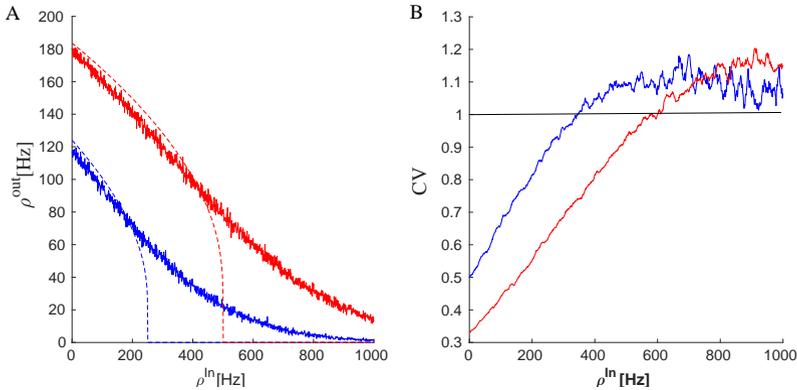}
		
		\caption[ Output firing and interspikes interval CV
                vs. Poisson input rates.]{(A) Firing rates of a neuron
                  receiving  excitatory Poisson input with
                  two  different excitatory
                  rates (the red curve corresponding to the higher one) vs. the
                  Poisson inhibitory input. Dashed lines are the response of
                  the neuron to the constant input with a magnitude equal to
                  the  Poisson rates (Equation \ref{eq:cteR}). (B) Coefficient of variation of the
                  spike intervals of a neuron receiving Poisson inputs of the
                  same rates as in the left graph. Near cutoff, the  neuron fires with CV close to one.} \label{fig:1}
	\end{figure}

If input rates satisfy Equation \ref{eq:Cond}, the output firing rate will be  
\begin{align}
\rho _{out} =   (g_{leak} + g_{exc}^0 w_E \rho_E \tau+ g_{inh}^0 w_I \rho_I \tau)  * (\log \dfrac{V_{rest} - V_{st}}{ V_{th} - V_{st} } )^{-1}	\label{eq:cteR}
\end{align}	
 The left-dashed curves in Figure \ref{fig:1} show the output firing rate for three different values of excitatory input rate versus inhibitory input rate.	
In the rest of this section we take the input to the neuron as stationary homogeneous Poissonian inhibitory and excitatory spike trains. In this case the number of spikes in a time interval $\Delta t$ follows a Poisson distribution:
\begin{align}
p(k_{[t,t+\Delta T]}) = (\lambda \Delta T)^k \dfrac{e^{ -\lambda  \Delta T }}{k!}
\end{align}
The output firing rate of the neuron to the Poisson input is depicted in
Figure \ref{fig:1}A. Compared to the constant input with the same constant rate as the
Poisson rate  $\lambda\ $, the curve becomes smoother and the transition from
silent state to active state does not show a sharp jump. Below the critical
inhibition value, the neuron output follows the mean-field deterministic
trajectory, however close to this point the fluctuation effect caused by
stochastic arrival of spikes manifests itself. Moreover, the stochasticity in
the input leads to stochastic firing at the output. Figure \ref{fig:1}B shows how the
coefficient of variation of the firing time interval of the output spike train
change according to the input. This quantity is calculated as
\begin{align}
CV(\delta t ) = \dfrac { \sigma _{\delta t}}{\langle \delta t \rangle}
\end{align}	
where $\delta t$ is the set of firing time intervals of the response of the
target neuron subjected to a stationary Poisson input. When the excitatory input is much stronger than the inhibitory one the output firing pattern becomes more regular and the CV value is small. However, close to the inhibition cutoff, CV  becomes close to unity, which is characteristic of the Poisson point process. 

	The Poisson input in the limit of a high firing rate and small
        synaptic weights can be approximated by a diffusion process. Suppose,
        in the time interval $[t,t+dt]$, $N(t , t+dt)$ excitatory spikes
        arrive at the cell each with synaptic strength $w_e$. As the spike
        arrival is a Poisson process with the rate $\lambda$ the distribution
        of $N(t , t+dt)$ is Poisson and all the cumulants of the random
        variable $N$ are equal to $\lambda dt$. This leads to the following cumulant for $ I(t,t+dt)=w_e N(t , t+dt) $ : 
\begin{align}
	\kappa _1 = & \langle I_{t,t+dt} \rangle dt = w_e \lambda dt \nonumber \\
	\kappa _2 = & Var(I_{t,t+dt} )  = w_e^2 \lambda dt \nonumber \\
	\kappa _3 = &  w_e^3 \lambda dt
	\end{align}
Higher cumulants can be ignored if we assume that $w_e^3 \lambda$ goes to zero
in the limit of a high number of afferent inputs. In theory, this can be
achieved by assuming weights to scale as $w_e = \dfrac{W}{\sqrt{k}}$ where $k$
is the number of presynaptic neurons. In this case, $\lambda \sim O(k)$ and the average excitatory and inhibitory currents are each of order $O(\sqrt{k})$, the variance of the current is of $O(1)$ and higher cumulants vanish in the limit of large $k$. 
In this case, one can take $ I(t,t+dt)$ as a Gaussian random variable with
mean and variance given by the above equation. It can also be written as 
\begin{align}
 I(t)dt = w_e \lambda dt + w_e \sqrt{\lambda} dW_t
\end{align}	
where $W_t$ is a Wiener process. 

In the conductance based model the input to the cell causes a change in the
conductance. As the input is stochastic the conductance is also a stochastic variable which can be written as 
  \begin{align}
	g(t)=\int_{-\infty}^{t} g_0 e^{-\dfrac{t-s}{\tau}} I(s)ds  = \int_{-\infty}^{t} e^{-\dfrac{t-s}{\tau}} g_0 w_e \lambda ds 
	+ \int_{-\infty}^{t} e^{-\dfrac{t-s}{\tau}} g_0 w_e \sqrt{\lambda} dW_s
	\end{align}	
The second term is the integral of a Wiener process with exponential
kernel. Fora  stochastic process  $Y_t = \int_{-\inf}^{t} f(s)dW_s$, we can easily verify that:
\begin{align}
 \langle Y_t \rangle = & 0 \nonumber \\
 \langle Y_t^2 \rangle = & \int_{-\inf}^{t} f(s)^2 ds
\end{align}	
If we consider a stationary and homogeneous Poisson process as the input then $g(t)$ will also attain a stationary probability distribution.
Using the above equation, mean and variance of $g(t)$  reach the  limits
\begin{align} 
	\langle g \rangle = & \tau g_0 w_e \lambda   \nonumber \\
	Var(g) = & \dfrac {\tau g_0^2 w_e^2 \lambda}{2} \label{eq:var}
	\end{align}	
  \begin{figure}
			\centering
			\includegraphics[width=1\linewidth,trim={0cm, 0cm, 0cm, 0cm},clip]{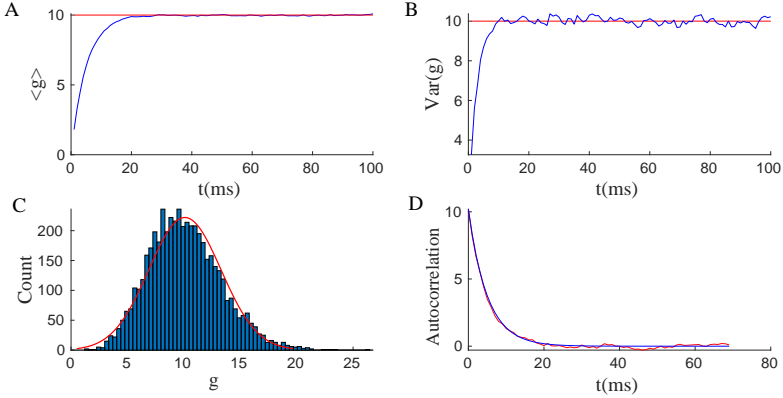}
		
		\caption[Diffusion Approximation of Poisson Input]{ Mean (A), Variance (B), stationary
                  probability distribution (C), and auto-correlation (D) of the
                  conductance $g(t)$ of a neuron receiving Poisson input. The
                  red lines are the values calculated by the diffusion
                  approximation and the red curve in the bottom left plot is
                  the Gaussian distribution with the mean and variance as
                  derived in the text. The auto-correlation matches Equation
                  \ref{eq:corr}. This figure shows that the  diffusion approximation is valid. } \label{fig:2}
	\end{figure}	
	With the above scaling of the weights, higher cumulants  vanish and
        one can assume that $g(t)$  reaches a stationary Gaussian probability
        distribution with mean and variance as above. The covariance of
        this process can be derived by direct multiplication and averaging
        over the noise terms of $g(t)$ and $g(t')$ from Equation \ref{eq:var} to reach the following equation when $t'>t$: 
\begin{align} 
 \langle g(t)g(t')\rangle - \langle g(t)\rangle \langle g(t')\rangle = Var(g) e^{ - \dfrac{(t'-t)}{\tau}} \label{eq:corr}
\end{align} 
The same procedure applies to inhibitory currents. Figure \ref{fig:2} shows statistics of conductance $g(t)$ and the validity of diffusion approximation. 
If we assume that the synaptic time scale $\tau$ is very small in comparison
to the membrane potential time scale, then one can ignore the cross
correlation and consider $Y_t = g(t) - \langle g \rangle $ as Gaussian white
noise. In this limit, 
\begin{align}
 \lim_{\tau \to 0} \langle Y_t Y_s \rangle  = 2 \tau var(g)  \lim_{\tau \to 0} \dfrac {e^{ - \dfrac{(s-t)}{\tau}}}{2\tau}  =  2 \tau var(g)  \delta (t-s)  
\end{align}	
 This leads  to a stochastic differential equation for the membrane potential evolution in the conductance-based model, when $v(t) < V_{th}$: 
  \begin{align} 
	C \dfrac{d v(t)}{dt} = & [g_{Leak}( v_{Leak} - v(t)) 
	 g_0 w_e \lambda _e \tau ( v_{Rexc} - v(t))   g_0 w_i \lambda _i \tau ( v_{Rinh} - v(t)) ] \nonumber \\
	 & +  \xi _{exc}(t) ( v_{Rexc} - v(t)) + \xi _{inh}(t) ( v(t) - v_{Rinh}) \nonumber \\ 
	 & \equiv  [a-bv] + \xi _{exc}(t) ( v_{Rexc} - v(t))  + \xi _{inh}(t) ( v(t) - v_{Rinh}) \label{eq:Gauss}
\end{align}
	
	Here $\xi _{exc}(t)$ and 	$\xi _{inh}(t)$ are purely random
        Gaussian processes:
\begin{align*}
	&\langle \xi _{exc}(t) \rangle = \langle \xi _{inh}(t)\rangle =0 \\
	& \langle \xi _{exc}(t)\xi _{exc}(t') \rangle =  \tau ^2 g_0^2 w_e^2 \lambda _e \delta (t-t') \equiv D_e \delta (t-t')\\
	& \langle \xi _{inh}(t)\xi _{inh}(t') \rangle =  \tau ^2 g_0^2 w_i^2 \lambda _i \delta (t-t') \equiv D_i \delta (t-t')	.
	\end{align*}	
	
\begin{figure}
\includegraphics[width=1\linewidth,trim={0cm, 0cm, 0cm, 0cm},clip]{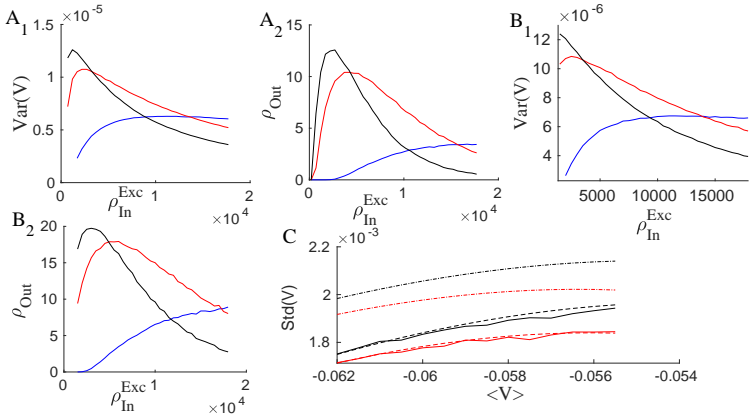}\\

		\caption[Stationary Membrane Potential Variance]{(A1-A2) Stationary variance of the membrane potential and
                  the output firing rates, when the excitatory rate ($x$-axis)
                  and inhibitory rates are balanced so that the average membrane potential is at $-57mv$ for three different values of $\tau_{syn} = [1$ (Blue), $3$ (Red), $5$ (Black)] ms. (B1-B2) Same with average membrane potential tuned at the threshold value $-50mv$. (C)
Stationary potential variance for a neuron receiving two different excitatory input rates, 1000Hz(red) and 1600Hz (blue), and corresponding inhibitory input, which places the average potential at a specific value shown in the $x$-axis. Solid lines are simulation results, dashed lines correspond to low firing regimes with tau approximation (Equation \ref{eq:A2.1}), and fine dashed lines show Gaussian approximation. ($\tau_{syn}=2.5ms$) } \label{fig:3}
\end{figure}

The first line of Equation \ref{eq:Gauss} is the deterministic evolution of the
potential. When the fixed point of the deterministic term, $v_{det}^{\inf} =
V_{st} = \dfrac{a}{b} $ as defined by Equation \ref{eq:Gauss} is greater than $V_{th}$,
the effect of fluctuations is marginal and the firing of the neuron is
governed by the drift term. However, when  $V_{st}$ is below the threshold,
fluctuations in the input can result in firing of the neuron. In Appendix \ref{app:1},
we calculate the variance and the mean of the membrane potential and the output firing rate of the neuron in the Gaussian approximation.
 We can improve the approximation for the potential distribution and the
 firing rate by considering the autocorrelation in the conductance. When
 $\tau$ is not negligible but small, we can use the $\tau$ expansion method to
 account for first-order corrections to the Fokker-Planck equation (Appendix
 \ref{app:2}). Figure \ref{fig:3}C shows that these corrections lead to a better approximation
 of the stationary membrane potential variance in a low firing rate
 regime. The stationary variance of the membrane potential decreases at higher values of $\tau_{syn}$ (Figures \ref{fig:3}A and \ref{fig:3}B) . Higher rates in excitatory and inhibitory input would lead to lower stationary variance and lower rates(Appendix \ref{app:3}).

\subsubsection{Condition for Poisson output} \label{sec3-sub1-1}

In the model of conductance based synaptic currents, the output spike train is
Poisson in the regime of fluctuation driven spiking when the stationary
average potential is away from the threshhold. To analyze the condition for
Poisson firing, we investigate the case where the stationary membrane
potential is below firing threshold. When excitatory and inhibitory currents
are matched in a way that the stationary membrane potential $V_{st}$ is close
to the firing threshold $V_{th}$, the approximations in the previous section
are no longer valid. 
In the limit of a high number of affarents the drift term would set the average membrane potential to its stationary value in a short time. Here we calculate the mean firing time and variance of it based on the approximation that fluctuations in the input close to the stationary average potential are weakly dependent on the voltage level.
 
 Therefore, our problem is reduced to the well known problem of the first
 passage time of a Brownian particle evolving as $ \dfrac{dx}{dt} = -kx + \xi
 (t)$ to reach the threshold $a$.  $\xi$ is white noise with variance
 $\sigma$. We want to approximate results for first passage time moments when
 the stationary membrane potential is close to the threshold. The goal here is
 to obtain approximate analytical results for firing rate and CV of the spike time interval to identify conditions for output Poisson firing and linearization of the output rate.

 We apply the Siegert formula (\cite{Siegert51}) for the first passage time
 (Appendix \ref{app:4}) in our case. Let us take $ x = v - \dfrac{a}{b}$,  $x_0 = v_0 -
 \dfrac{a}{b}$ and $x_{th} = v_{th} - \dfrac{a}{b}$ where $\langle
 V\rangle_{st}=\dfrac{a}{b}$. Thus, the  random variable $x$ evolves as $ \dfrac{dx}{dt} = -bx + \sigma(x) \xi (t)$. where $\xi$ is  a white noise with unit variance and 
\begin{align}
&\sigma^2(x) =  \dfrac{1}{C^2}\{ D_e( v_{Rexc} - \dfrac{a}{b} +x)^2   +  D_i( v_{Rinh} - \dfrac{a}{b} +x )^2 )\} \nonumber\\
 &b = \dfrac{1}{C} (g_{Leak} + g_0 w_e \lambda _e \tau + g_0 w_i \lambda _i \tau )
\end{align}
For the case of Poisson current to the cell, if the mean value is close to the threshold,  the magnitude of fluctuations does not vary much in the interval $[V_{rest} ,V_{th}]$ , therefore in the following we neglect dependence of $\sigma$ on $x$.

The approximation for the mean passage time from the Siegert formula is
\begin{align} 
t_1(x_{th} \mid x_0)  = & -\dfrac{\sqrt{\pi}}{b}\kappa + [(\dfrac{a}{b} - V_{rest})\sqrt{\dfrac{b}{\sigma^2}} ] \dfrac{\sqrt{\pi}}{b} +  \dfrac{\sqrt{\pi}}{b} [ (V_{th} - \dfrac{a}{b})]  \sqrt{\dfrac{b}{\sigma^2}} \nonumber \\ & + \dfrac{1}{b} [ (V_{th} - \dfrac{a}{b}) \sqrt{\dfrac{b}{\sigma^2}} ]^2 \label{eq:fpt}
\end{align}
where  $\kappa$ is very weakly dependent on input rates and we take it as a
constant factor (see Eq.\ref{eq:App4} in Appendix\ref{app:4}). Altogether, we can write down the
average rate of firing, $\rho= \dfrac{1}{t_1(x_{th} \mid x_0)}$, in the case that the
average stationary potential is close to threshold as 

\begin{align}
\rho = \dfrac{b}{\sqrt{\pi}} (\dfrac{1} { \kappa + (V_{th} - \dfrac{a}{b})  \sqrt{\dfrac{b}{\sigma^2}}}) \label{eq:rateAp}
\end{align}

\begin{figure}
		
			\centering
			\includegraphics[width=1\linewidth,trim={0cm, 0cm, 0cm, 0cm},clip]{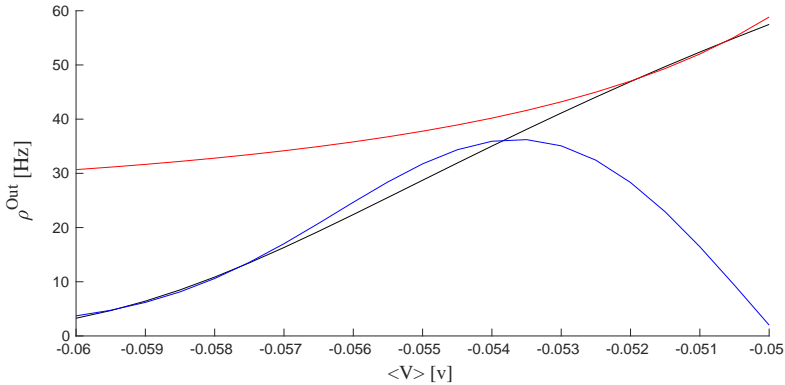}
		
		\caption[Output firing rate using Gaussian approximation.]
                {Firing rate approximation by Gaussian assumption of Eq.\ref{eq:App1} in  Appendix\ref{app:1} (blue) and by near-threshold high firing
                  assumption(red) of Eq.\ref{eq:rateAp} are compared with simulation
                  results (black curve). Here, we fix the inhibitory input and
                  select a value of the  excitatory input rate that leads to a specific mean stationary membrane potential shown on the $x$-axis. } \label{fig:4}
	\end{figure}

In Figure \ref{fig:4}, we have plotted this rate approximation which causes only a
  very small error when the average potential is near threshold. If $\langle
v \rangle = \dfrac{a}{b}$ is constant for balanced inhibitory and excitatory
input rates, then the output rate of the neuron would be linearly proportional
to the input rates via the factor $b$. Moreover the output rate would decrease
as $\dfrac{1}{x_{th}}$ with the increase of the distance of the stationary average potential from the threshold.

 Next, we want to investigate the variance of first passage time. From the
 recursion formula (Eq.\ref{eq:app4-1} in Appendix\ref{app:4}) and after proper approximations we end up
 with the  following expression for the CV of the time interval between spikes: 
\begin{align}
CV^2 = \dfrac{Var(t)}{\langle t \rangle ^2} \approx 1 + \dfrac{C}{t_1(x_{th} \mid x_0)^2} + 2\sqrt{\pi}ln(2) \dfrac{\dfrac{ x_{th} }{b\sqrt{b} \sigma}}{t_1(x_{th} \mid x_0)^2 } 
\end{align}
where $C$ is a negative constant (see Eq.\ref{eq:app4-4} in Appendix\ref{app:4} ).The second term is  negative
and  monotonically goes to zero as $x_{th} $ increases. In the limit of large
$x_{th}$,  the second and the third term both go to zero and $CV$ approaches
$1$. However, in the near threshold approximation the  maximum of the third term occurs where  CV is approaching $1$. Expanding in  powers of $x_{th}$ , we arrive at 
 
\begin{align}
x_{th}^{opt} := V_{th} - \langle V \rangle_{st} = \dfrac{\pi \sigma}{2 \sqrt{b}} \label{eq:CV}
\end{align}

\begin{figure}
			\centering
			\includegraphics[width=1\linewidth,trim={0cm, 0cm, 0cm, 0cm},clip]{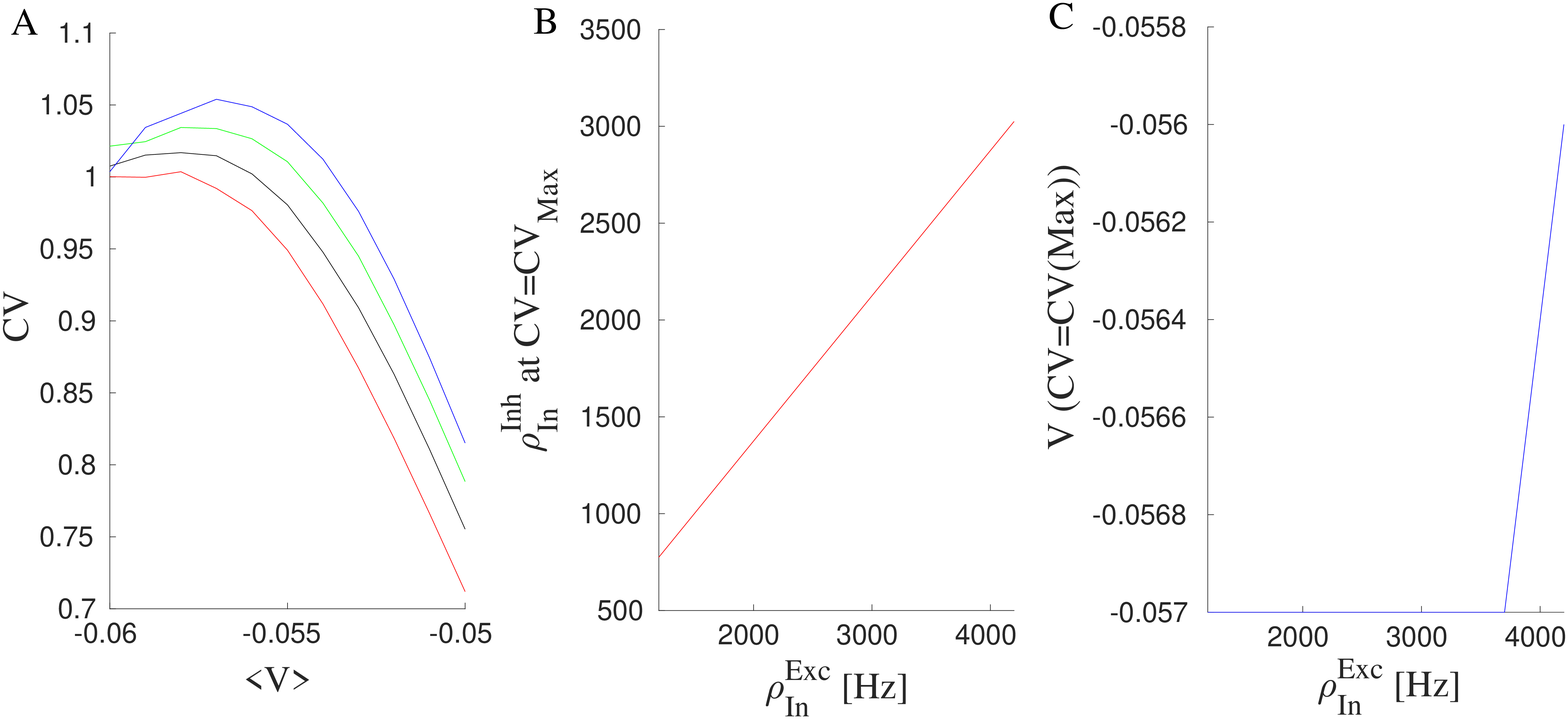}
		
		\caption[CV of Interspike Intervals]{(A) CV of interspike
                  intervals  for four different excitatory input rates and
                  their corresponding inhibitory rates, which set the average
                  membrane potential at each specified value shown in the
                  x-axis.(Red curve corresponds to the highest excitatory rate (4000Hz) and the blue one to the lowest rate (1000Hz) ) (B) Inhibitory rate vs. excitatory rate at the
                  maximal $CV$. (C) Membrane potential value at the value of the maximal $CV$. } \label{fig:5}
              \end{figure}

As shown in Appendix\ref{app:3} (Eq.\ref{eq:app3-1}), $\dfrac{\sigma}{\sqrt{b}}$ reaches a constant value for high input rates. This can be used to determine the value of $ \langle V \rangle_{st}$ that leads to maximal CV. 
Figure \ref{fig:5} shows the CV of the interspike interval for different sets of excitatory
and inhibitory pairs of input. As can be seen, at the threshold, neuronal
firing time intervals have lower variance, but the CV approaches 
one far away from the threshold. The stationary membrane potential value
corresponding to the maximal value of CV from  Equation \ref{eq:CV} is shown in the
right diagram and it matches well with the actual values from the
simulation. At $V_{P}:= \langle V \rangle_{st}^{opt} \approx -0.56 mv$, the CV
for different input rates has a maximum  independently of the rate  values.

  In the middle plot, we see that the inhibitory rate which satisfies
  $CV=CV_{max}$ varies linearly with the excitatory rates. As can be seen,
  when the stationary membrane potential is approximately below $V_{P}$, the
  CV of interspike intervals approaches $1$, independently of the values of inhibitory and excitatory rates.
This is an indicator that output firing in response to Poisson input is itself
a Poisson point process when $\langle V \rangle_{st}$ lies below $V_P$ . For a
more conclusive result, one has to calculate higher moments or investigate the
limit of the FPT probability density when $x_{th}$ is very large.
The fact that the Poisson output condition for different sets of Poisson input
leads to approximately a similar level of the membrane potential  enables us
to introduce the linearization of the output rate at the line corresponding to
$\langle V_{m}\rangle = V_p$.

\subsubsection{Linear Poisson neuron approximation} \label{sec:linear}

\begin{figure}
			\centering
			\includegraphics[width=1\linewidth,trim={0cm, 0cm, 0cm, 0cm},clip]{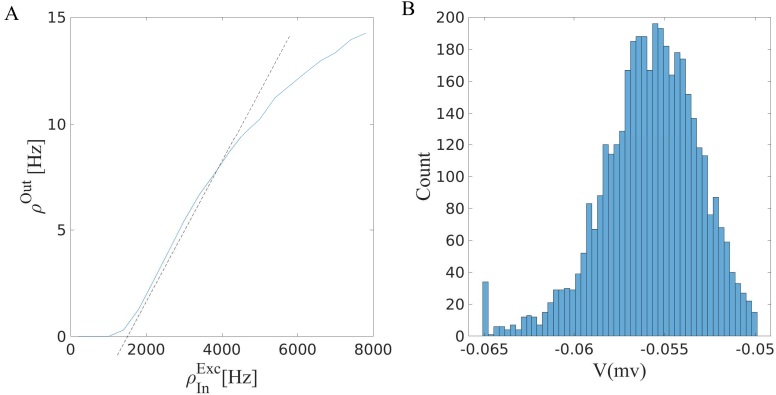}
		
		\caption[Response of the neuron to the balanced current near
                the threshold.]{Response of a population of neurons receiving
                  excitatory and inhibitory inputs balanced in a way that the
                  drift term has a fixed point at $V_P = -0.56mv$.  (A)
                  Output firing rate for different values of balanced
                  inhibitory and excitatory input rates. The output rate
                  changes semi-linearly on this line and firing in this regime
                  that is driven by the fluctuation in the input causes the
                  neuron to fire with Poisson point process
                  statistics. (B)The stationary potential distribution of
                  the population of neurons. There is a reservoir of neurons close to the threshold while the average firing rate is about 20 Hz. Parameters used: $w_E$ =0.5 ,$ w_I= 0.75$ , $N_E= 7000$ , $N_I=0.25* N_E$ } \label{fig:6}
	\end{figure}

Here we want to show that linearizing the response curve of a neuron receiving
Poisson current near $V_P$, introduced in the last subsection,  leads to a
good approximation for the firing rate of the neuron in a wide range of input
rates. The linearization is around the line characterized by Equation \ref{eq:VST} 
with $V_{st}= V_{P}$ in the  $\rho _{exc} - \rho _{inh}$ plane. This line
corresponds to the balance of mean excitation and inhibition at $V_{P}$. On
this balance line from Equation \ref{eq:rateAp}, the output rate will  depend linearly on the excitatory or inhibitory input rate (see Figure \ref{fig:6}).

We want to linearize the output rate around $V_P$ . For this purpose let us
write the equation of the plane passing through the line of current balance at $V_P$ (Eq.\ref{eq:Bal}) and the tangent line in the $(\rho_E , \rho_{out} )$ plane at some arbitary point $( \rho_{I}^0 , \rho_{E}^0 , \rho_{out}^0)$. The balance condition line for an excitatory neuron connected to $k_{EE}$ excitatory neurons and $k_{IE}$ inhibitory neurons each firing with the rate $\rho_E$ and $\rho_I$, respectively, and receiving external excitatory rate $\rho_{Ext}$ is of the  form:
\begin{align}
\rho _E^e * k_{EE} = &  \dfrac{(V_{Rinh} - V_{P})*g_{inh}^0*w_{EI}}{ g_{exc}^0*w_{EE}*(V_{P}- V_{Rexc})} \rho_I *k_{EI} \nonumber \\ &+  \dfrac{g_{leak}(V_{rest} - V_{P}) }{\tau* g_{exc}^0*w_{EE}*(V_{P}-V_{Rexc})} -  \dfrac{\rho _{Ext}^e}{w_{EE}} \label{eq:Bal}
\end{align}
  We rewrite this in simpler form as $\rho _E^e = k \rho_I  +C$.   The
  equations for the balance line and the other tangent line in the $(\rho_E , \rho_{out} )$ plane  are
  
\begin{align} 
&\dfrac{ (\rho _E - \rho_E^0)  }{k} =  \rho_I - \rho_I^0 = \dfrac{ \rho _{out} - \rho_{out}^0 }{\alpha_{OI}} \nonumber \\
& \dfrac{ \rho _O - \rho_O^0 }{\beta_{OE}} =  \rho_E - \rho_E^0
\end{align}

Therefore, the equation of the plane passing through these lines is of the form

\begin{align}
( \rho _{out} - \rho_{out}^0 ) = & \beta_{OE} ( \rho_E - \rho_E^0)  + ( \alpha_{OI}-\beta_{OE}k )(\rho_I - \rho_I^0)  \label{eq:plane}
\end{align}

\begin{figure}
			\centering
			\includegraphics[width=1\linewidth,trim={0cm, 0cm, 0cm, 0cm},clip]{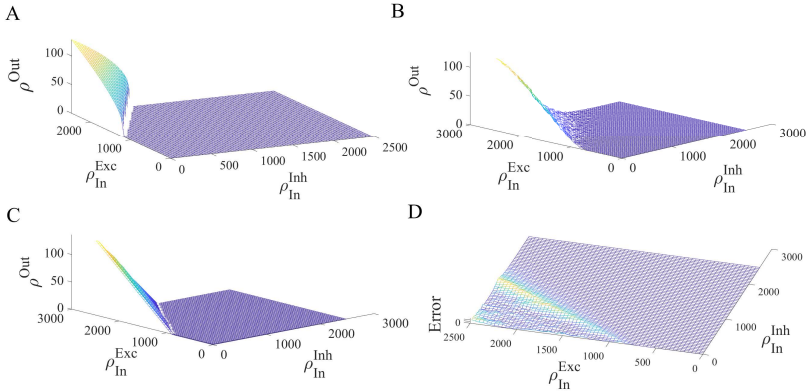}
		
		\caption[Linear Poisson Neuron approximation.]{(A)
                  Firing rate of a neuron w.r.t. different values of constant
                  inhibitory and excitatory input. (B) The same for
                  Poisson input. (C) The linear approximation for
                  the output on the critical line of Eq.\ref{eq:Bal}. (D) The error of the linear Poisson neuron approximation.} \label{fig:7}
	\end{figure}

$\beta_{OE}$ is the derivative of the nonlinear response at the selected point
in the  direction of  $\rho_E$, and $\alpha_{OI}$ is proportional to the
change of output rate by changing inhibition and accordingly excitation on the
balance line. These derivatives do not vary much on the balance line,
therefore, the choice of the linearization point does not matter for us at
this stage. This suggests that the plane of Equation \ref{eq:plane}  is tangent to the
$\rho_{out}$ surface. This linear approximation, however, fails for very high
excitatory input where  the saturation of the neuron causes non-linearity. The
linearization point is where the output firing curve has the lowest curvature,
and therefore the second derivative vanishes, which makes the approximation error minimal. Figure \ref{fig:7} shows the output firing rate of the target neuron and the linear approximation presented above.

In the next section, we want to investigate the homogeneous firing state of a
network. For this purpose we will look at self consistency solutions $
\rho_{out} = \rho_E(in) = \rho_E^*$  for an arbitrary value of inhibitory
current. From Equation \ref{eq:plane} :

\begin{align}
 (1 - \beta _{OE}) \rho_E^* =  ( \alpha_{OI}-\beta_{OE}k )(\rho_I - \rho_I^0) + \rho_O^0 - \beta_{OE}\rho_E^0 
\end{align}

Putting in $ k\rho _I^0 - \rho_E^0 = -C $ and dividing the above equation by $\beta_{OE}$, we arrive at 

\begin{align}
 (\dfrac{1}{\beta_{OE}} - 1 ) \rho_E^* =   - k\rho_I -C  +   \dfrac {\alpha_{OI}}{\beta_{OE}} (\rho_I - \rho_I^0 )  +  \dfrac {1}{\beta_{OE}}\rho_O^0
\end{align}

$\beta_{OE}$  depends on the number of excitatory input to the cell, $K_{EE}$, and is related to the proportional change of output firing at the balance line to the change in the firing rate in each excitatory neuron. On the other hand, $\alpha_{OI}$, proportional to change in the firing rate while fixing the balance condition, is much smaller than  $\beta_{OE}$. Therefore, when $K_{EE}$ is large, the self-consistency equation matches the balance line of Equation \ref{eq:Bal} with a minimal error. 

\subsection{ Sparse homogeneous EI population dynamics  }

As we have seen for sets of Poisson input that produce low firing output the
statistics of spiking events resembles a Poisson process. In a population of
neurons there might be a stable stationary or oscillatory population rate with
Poisson firing  of individual neurons. In this case, the magnitude of
fluctuations in the population average scales as $O(N)$. This inhomogeneous
synchronous or asynchronous firing state exists only in the low firing
regime. In the high firing state, the large imbalance of excitatory and inhibitory input leads to periodic firing of the individual neurons which can be also synchronized with high amplitude and high frequency oscillatory population rates. 
We can use the linear Poisson approximation for identifying and analyzing the dynamics in the low firing rate regime which is of most interest to us.
In a homogenous population, solutions of the self-consistency equations for both inhibitory and excitatory neurons' average output firing rate receiving synaptic currents originated from both neurons in the population  and external inhibitory and excitatory currents $\lambda$ can be written as follows:
\begin{align}
\rho_{E}^{st} &= f( k_{EE}\rho_{E}^{st} , k_{EI} \rho_{I}^{st} ,\lambda_{EE} , \lambda_{EI} )  \label{eq:MFR1}
\end{align}
\begin{align}
\rho_{I}^{st} &= g( k_{EI}\rho_{E}^{st} , k_{II} \rho_{I}^{st} ,\lambda_{IE} , \lambda_{II}  ) \label{eq:MFR2}
\end{align}	
for $ \rho_{E} , \rho_{I} \in [0, \rho_{max}  ]	$ .	Functions $f$ and $g$ are called excitatory and inhibitory gain functions and $k_{xy}$ is the number of internal connections between neurons in the population.  Solving for these gain functions in the general case is not analytically
tractable for the EI population. Dynamics to the stationary rates given by equations  \ref{eq:MFR1}-\ref{eq:MFR2} can be phenomenologically approximated by the following mean field equations:

\begin{align} 
\dfrac{d\rho_{E}}{dt} &= - \dfrac{1}{\tau_m}(\rho_E(t) - f( k_{EE}\rho_{E}(t) , k_{EI} \rho_{I}(t) ,\lambda_{EE} , \lambda_{EI} )) \nonumber\\
\dfrac{d\rho_{I}}{dt} &= -\dfrac{1}{\tau_m}(\rho_I(t) - g( k_{EI}\rho_{E}(t) , k_{II} \rho_{I}(t) ,\lambda_{IE} , \lambda_{II} ) ) \label{eq:DynamicSystem}
\end{align}	

This set of equations may have multiple
solutions and changing control parameters can lead to Hopf and saddle-node
bifurcations, which in turn produce/destroy oscillations or produce/destroy
pairs of fixed points.
Although it is possible to numerically investigate the FPE for EI populations
and its bifurcation diagram, in the next subsections, we follow another
approach by using linearized nullclines approximation and logistic function
approximation for functions $f$ and $g$. We show that studying these model systems is appropriate for the bifurcation analysis and agrees with simulation results.

\subsubsection{Linearized nullclines}

\begin{figure}
			\centering
			\includegraphics[width=0.8\linewidth,trim={0cm, 0cm, 0cm, 0cm},clip]{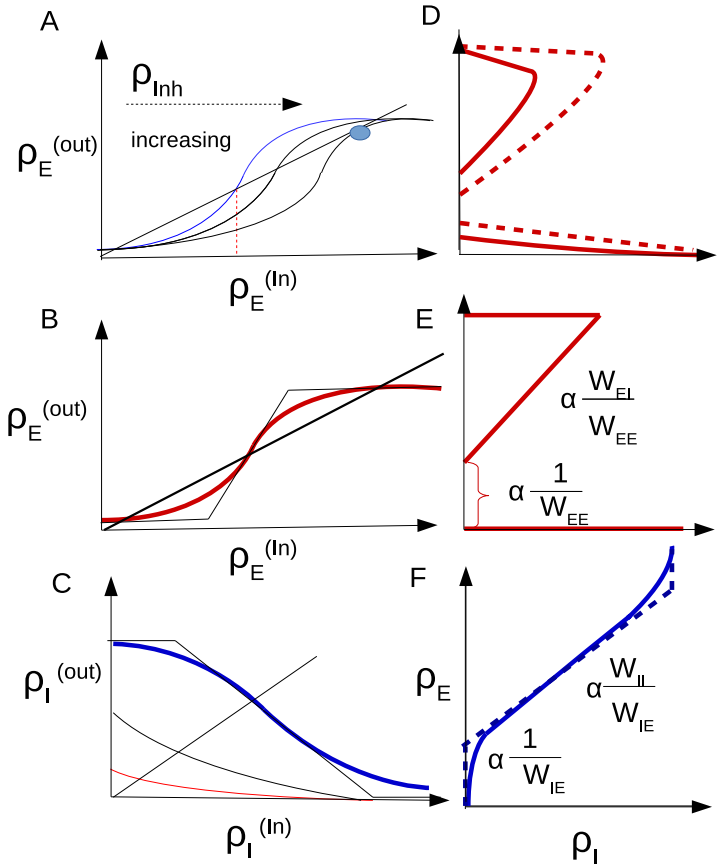}
		
		\caption[Linearized gain functions and nullclines.]{(A)
                  Excitatory neuron output rate vs. excitatory input rate at
                  three fixed values of inhibitory currents. (B)
                  Linearized excitatory gain function. (C)
                  Inhibitory neuron output rate vs. inhibitory input at three
                  different values of excitatory current. (D) Excitatory nullclines of
                  Equations \ref{eq:MFR1}-\ref{eq:MFR2} for two different values of $w_{EE}$ with the dotted curve coressponding to the higher value. 
       (E) Linearization of the excitatory nullcline (F)  Inhibitory nullcline and its linearization based on Equations \ref{eq:line1}-\ref{eq:line2} (dotted curve).} \label{fig:8}
	\end{figure}

Function $f$ in the Equation \ref{eq:MFR1} for the stationary excitatory rate is of the form of an S-shape or sigmoidal curve. Therefore, this equation has one or three solutions depending on
the value of the inhibitory rate. This is shown in Figure \ref{fig:8}A for three
different total inhibitory currents.  For a low to a moderate value of
inhibition there exist three fixed points, i.e. intersections of the linear
line with the sigmoidal gain function, at quiescent state, semi-linear
section, and high firing state. Increasing the inhibitory input rate causes
the nonlinear gain function to move to the right to the point specified in the
graph by a blue dot, and eventually the middle saddle and high fixed point
annihilate each other through a saddle-node bifurcation. On the other hand,
increasing external excitatory input will move the graph upward, which leads
to the annihilation of the low fixed point and the saddle through another SN
bifurcation. Figure \ref{fig:8}D shows the solutions to the equation \ref{eq:MFR1} for a typical sigmoidal gain function and
different values of total inhibitory current to the excitatory
population. This is plotted for two different values of $w_{EE}$ with the
dashed curve corresponding to higher $w_{EE}$.

  Similarly, Figure \ref{fig:8}C is the plot corresponding to the Equation \ref{eq:MFR2}. Here, the nonlinear sigmoid function $g$ is plotted for three different values of excitatory current. There exists a single intersection between the line passing through the origin and these curves, which means Equation \ref{eq:MFR2} has a unique solution for the stationary inhibitory rate at each specific excitatory input. Figure \ref{fig:8}F is the plot of the location of these intersections for different values of inhibitory input.     
 As  can be seen in Figure \ref{fig:8}, there exists a semilinear section in the nullcline
 graphs corresponding to solutions in the linear Poisson section of gain
 functions. Based on the linear Poisson approximation of the section \ref{sec:linear}, the  equations for these lines in both excitatory and inhibitory nullcline graphs are: 

\begin{align}
\rho _E^{exc} * k_{EE} =  & \dfrac{(V_{Rinh} - V_{P})*g_{inh}^0*w_{EI}}{ g_{exc}^0*w_{EE}*V_{P}} \rho_I *k_{EI} \nonumber\\ & +  \dfrac{g_{leak}(V_{rest} - V_{P}) }{\tau* g_{exc}^0*w_{EE}*V_{P}} -  \dfrac{\lambda_{EE}}{w_{EE}} \label{eq:line1}
\end{align}

\begin{align}
\rho _E^{inh} * k_{IE} = & \dfrac{(V_{Rinh} - V_{P})*g_{inh}^0*w_{II}}{ g_{exc}^0*w_{IE}*V_{P}} \rho_I * k_{II} \nonumber \\ & +  \dfrac{g_{leak}(V_{rest} - V_{P}) }{\tau*g_{exc}^0*w_{IE}*V_{P}} - \dfrac{\lambda_{IE}}{w_{IE}} \label{eq:line2}
\end{align}
where $k_{\alpha \beta}$ is the number of excitatory/inhibitory synapse to an
excitatory/inhibitory neuron. In the remainder of this work, we assume
external inhibiotry currents to be zero, in line with our assumption that inhibition is local in our model. We assume  $\dfrac{k_{EI}}{k_{EE}} = \dfrac {k_{II}}{k_{IE}} $, which simplifies our analysis.

In the $\rho_{I} - \rho _{E} $ plane the slope and the $y$-intercept of
the 
two lines in Equations \ref{eq:line1}-\ref{eq:line2}  determine the intersection of the two nonlinear
nullclines and can be used to find  approximate locations of the 
bifurcation points of Equation \ref{eq:DynamicSystem}. We choose $\langle w_{EE}\rangle $ and
$\rho_{Ext}=\lambda _{EE}$ as control parameters of our model. Therefore, we
first discuss how their change  affects the nullclines of Equations \ref{eq:MFR1}-\ref{eq:MFR2}. Increasing $\rho_{Ext}$ moves the sigmoid graph in Figure \ref{fig:8}A
upwards causing the low and middle fixed points to move towards each
other. For a sufficiently high value of excitatory rate, these fixed points
will disappear by a saddle-node bifurcation. In the excitatory nullcline
graph (Figure \ref{fig:8}D) increasing  $\rho_{Ext}$ shifts the graph to the
right. Increasing $W_{EE}$ will both reduce the $y$-intercept of the
excitatory nullcline and the slope of the linear section as  shown in
Fig. \ref{fig:8}D. The nullcline for the inhibitory rate equation stays intact under change of control parameters.

The intersections of the inhibitory and excitatory nullclines are solutions of
the set of rate equations \ref{eq:MFR1}-\ref{eq:MFR2}. Based on the number of fixed points and their
stability, the  system can show  bi-stability of quiescent and high firing,
oscillatory dynamics, avalanches, high synchronized activity, and quiescent
state. Investigating the linearized sections of the graphs can help us
identify different regimes of activity. The slope and $y$-intercept of the
linear sections of both nullclines can be compared for this purpose. Based on
the Poisson neuron approximation there exists  a point in control parameter
space where the $y$-intercept and slope of two nullclines are equal. This
points  is the solution of the following linear constraints:

\begin{align}
 s_{exc} := \dfrac{ w_{EI}k_{EI}}{w_{EE}k_{EE}} = \dfrac{ w_{II}k_{II}}{w_{IE}k_{IE}} := s_{inh}
 \label{eq:slope}
\end{align}
\begin{align}
  y_{exc} := \dfrac{ d - \rho_{Ext}}{w_{EE}k_{EE}}  =  \dfrac{ d - \lambda_{IE}}{w_{IE}k_{IE}} := y_{inh} \label{eq:yinter}
\end{align}
where $d$ is a constant equal to $\dfrac{g_{leak}(V_{rest} - V_{th}) }{\tau* g_{exc}^0*(V_{th}-V_{Rexc})}$. 

\begin{figure}[t]
			\centering
			\includegraphics[width=0.8\linewidth,trim={0cm, 0cm, 0cm, 0cm},clip]{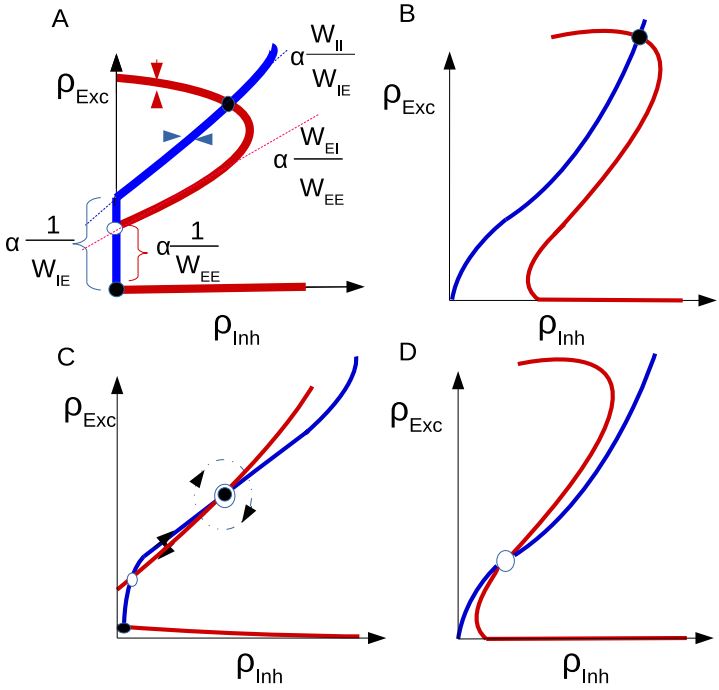}
		
		\caption[Nullclines' arrangments corresponding to different regimes of activity.]{ Nullcline diagrams corresponding to regimes of bistabilty (A), high synchronized firing(B), avalanches (C) and oscillatory dynamics (D). Red curves are excitatory nullclines (Eq.\ref{eq:MFR1}) and blue curves are inhibitory nullclines (Eq.\ref{eq:MFR2}).  } \label{fig:9}
	\end{figure}

 Figure \ref{fig:9}A shows the case in which $ w_{EE}w_{II} > w_{EI}w_{IE} $ and the
 $y$-intercept of the excitatory nullcline is lower than the inhibitory
 one. This occurs in the regime of a low to moderate imbalance of excitatory
 and inhibitory external input and high excitatory synaptic weight. In this
 case, the quiescent and the high firing fixed point are both stable and
 separated by a saddle. Increasing external excitatory input, the excitatory
 nullcline are shifted to the right and the middle saddle and quiescent node
 disappear by a saddle-node bifurcation and only the high firing synchronous
 state  remains (Figure \ref{fig:9}B). Increasing $w_{EE}$ has the same qualitative effect. However, decreasing external input or $w_{EE}$ drives the system to a quiescent state through different sets of bifurcations depending on the initial state of the system and in general on other parameters of the model. This intermediate transition state involves the appearance of a fixed point in the linear section.

When $s_{exc} > s_{inh}$ while $y_{exc} < y_{inh}$, there is a fixed point in
the linear section as depicted in Figure \ref{fig:9}C. We will discuss the stability of
the fixed point on the linear segment in the following sections. By increasing
external input, the quiescent fixed point and the low saddle move closer to
each other while the fixed point on the linear section ascends to higher rate
values. After the saddle-node bifurcation at the low rate, only the fixed point on the linear section survives as shown in Figure \ref{fig:9}D. These two arrangements when the fixed points are close to low firing regimes are important for us because of the avalanche dynamics that appear near this region.  
The intersection point of the nullclines in the semilinear regime can be
approximated by the intersection point of the linearized nullclines which is:
 \small
 \begin{align} 
&\rho _E^c = \dfrac{\tau g_0 (V^R_{E}-V_{th})(c_{II}\rho_{Ext} - c_{EI}\lambda_{IE}) + g_L(V_L - V_{th})(c_{II}-c_{EI})}{\tau( c_{IE}c_{EI}-c_{EE}c_{II})} \nonumber \\
&\rho _I^c =  \dfrac{\tau g_0(V^R_{E}-V_{th})(c_{EE}\lambda_{IE}-c_{IE}\rho_{Ext}) - g_L(V_L - V_{th})(c_{IE}-c_{EE})}{\tau(c_{IE}c_{EI}-c_{EE}c_{II})} \label{eq:InterPoint}
\end{align}
\normalsize
where $c_{xy} = k_{xy} w_{xy} g_{y}(V^R_{y}-V_{th}) $ .
 
\begin{figure}
			\centering
			\includegraphics[width=1\linewidth,trim={0cm, 0cm, 0cm, 0cm},clip]{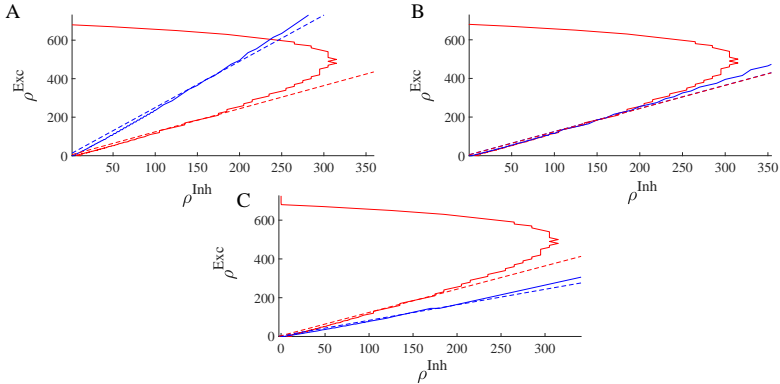}
		
		\caption[EI Nullclines in the case of Hopf bifurcation.]{Nullclines for excitatory and inhibitory neuron populations and their corresponding linear approximations of Equations \ref{eq:line1}-\ref{eq:line2} obtained from network simulation. Values of parameters are  $w_{EE}= [0.75$ (A)$,0.55$ (B)$,0.4$  (C)]$ , w_{EI}= 2 , w_{II} =1.5 , w_{IE} = 0.6 $.} \label{fig:10}
	\end{figure}

\begin{figure}
			\centering
			\includegraphics[width=1\linewidth,trim={0cm, 0cm, 0cm, 0cm},clip]{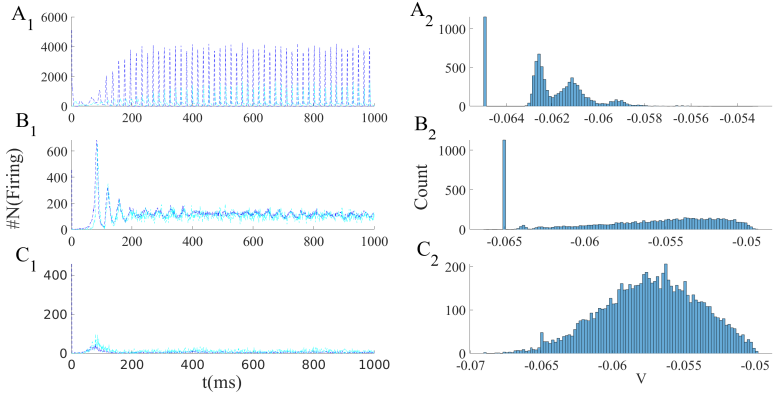}
		
		\caption[EI simulation at the oscillatory regime.]{Simulation
                  of populations of $N_E=10000$ excitatory and $N_I = 0.25N_E$
                  neurons connected by the same average synaptic weight values as in
                  Fig.10. (A1-B1-C1) Number of active excitatory neurons (dark
                  blue) and active inhibitory neurons (light blue) in each
                  time slot of (0.1ms) for three different values of
                  $w_{EE}$. (A2-B2-C2) The corresponding stationary membrane potential distribution. In the asynchronous state, the distribution has higher variance. } \label{fig:11}
	\end{figure}

 As discussed previously, in the intermediate parameter range, the  high fixed
 point might become unstable through either an Andronov-Hopf or a saddle node
 bifurcation. Figures \ref{fig:10}-\ref{fig:11} show nullcline graphs and population
 activity when the high fixed point loses stability by a Hopf
 bifurcation. Fig.\ref{fig:10}A shows nullclines of a system that has stable high and
 quiescent fixed points with a saddle node at low rates. By decreasing
 $w_{EE}$, $s_{exc}$  approaches $s_{inh}$, while sufficient external input
 guarantees that $y_{exc} <y_{inh}$ during this parameter change. In this
 particular setup, the inhibitory nullcline is semi-linear and we may
 speculate that the high fixed point goes through a Hopf bifurcation when the
 return point of the  excitatory nullcline touches the inhibitory nullcline
 which takes place at some value  $w^*_{EE} \in [0.55,0.75]$. Decreasing
 $w_{EE}$ further, the high saddle node descends through a linear segment and
 gets closer to the lower saddle point (Fig.\ref{fig:10}B). The  limit cycle becomes unstable by a
 saddle separatrix loop bifurcation. After saddle-node annihilation of
 low and high saddles,the system will end up in the quiescent state
 for low values of $w_{EE}$(Fig.\ref{fig:10}C). Population activity in these three
 regimes is shown in Fig.\ref{fig:11}. Neurons are firing synchronously at a high rates
 in three different sub-populations in the first case. The high oscillatory
 activity appears in the second regime where the unstable saddle, which is
 encircled by a stable limit cycle, lies close to the high activity
 region. The membrane potential distribution, in this case, has a higher variance, and neurons fire asynchronously.

\begin{figure}
			\centering
			\includegraphics[width=1\linewidth,trim={0cm, 0cm, 0cm, 0cm},clip]{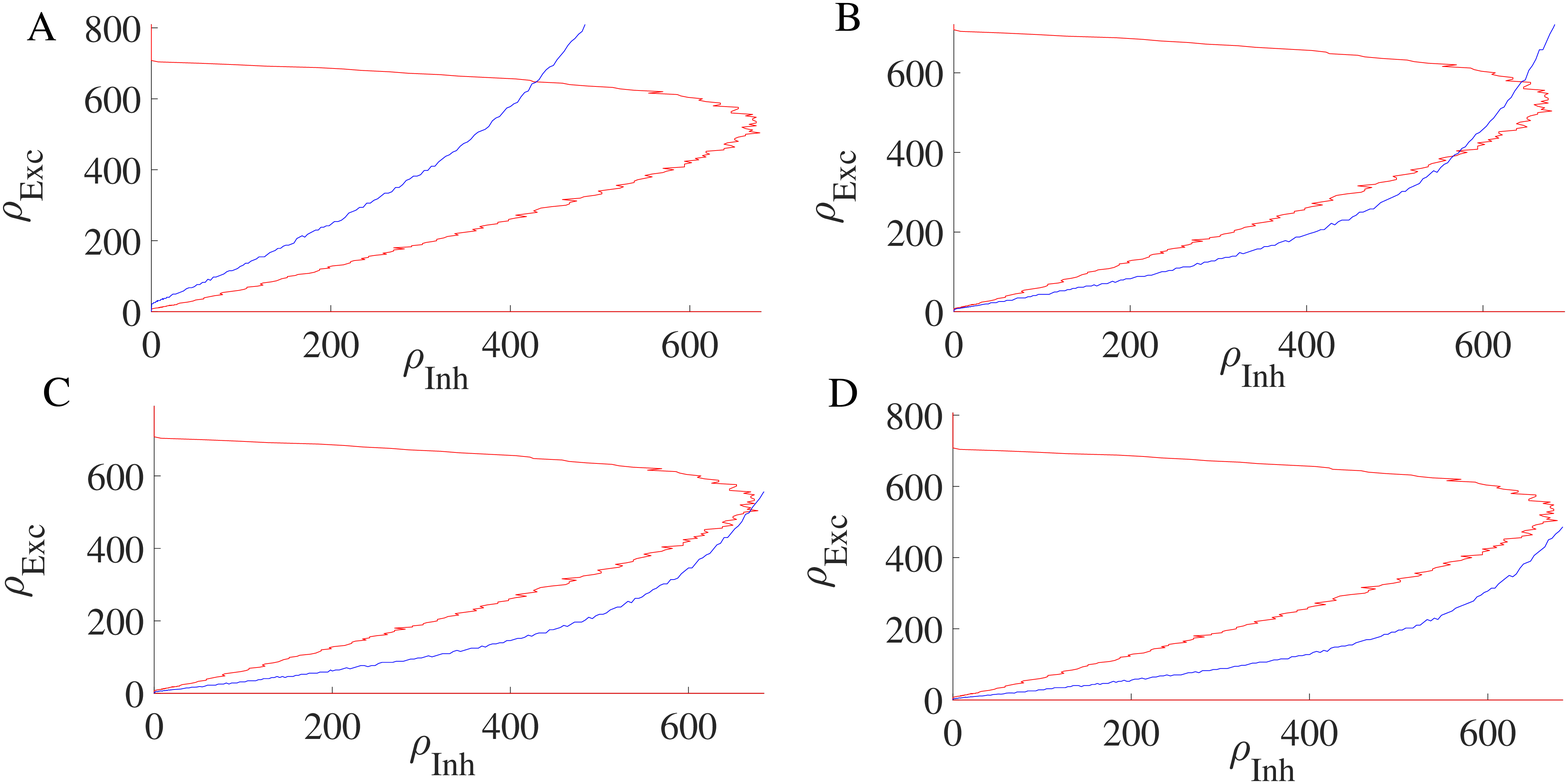}
		
		\caption[Nullclines in the saddle-node bifurcation
                case.]{Nullclines for excitatory (red curve) and inhibitory
                  (blue curve) population rates.The parameters used are
                  $w_{EE} = 0.6, w_{EI} = 1.2 , w_{II}= 0.6$ , $w_{IE} = 0.6 $
                  and  $w_{EE}= [0.8(A), 0.6(B), 0.4(C), 0.2(D)] $. Decreasing $w_{EE} $
                  changes the intersections of the  two curves.} \label{fig:12}
	\end{figure}

 On the other hand, Fig.\ref{fig:12} shows the case in which a high fixed point loses stability through colliding with the saddle that ascends along the linear section. This situation occurs at a lower level of external input, in which decreasing $W_{EE}$ causes $y_{exc}$ to pass above $y_{inh}$ before the slopes become equal. In this case, the high activity fixed point is annihilated by the saddle point.

\begin{figure}
			\centering
			\includegraphics[width=1\linewidth,trim={0cm, 0cm, 0cm, 0cm},clip]{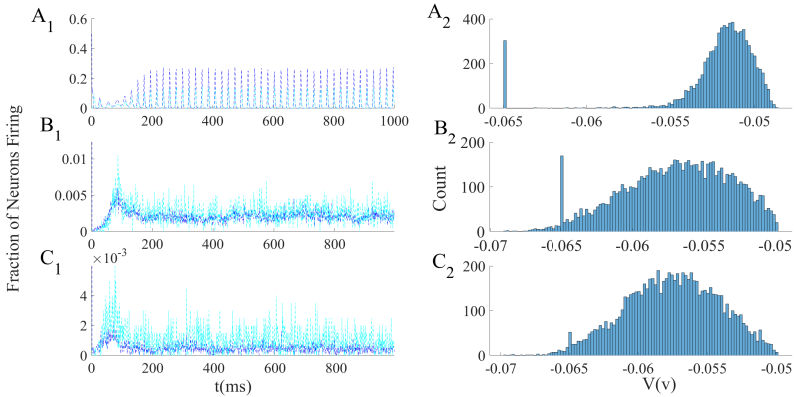}
		
		\caption[EI population: asynchronous state.] {Simulation
                  results of the network with same parameters as in Fig.\ref{fig:11}
                  except for  $ w_{II} =2.4 $. The EI population shows
                  asynchronous firing in the medium range of $w_{EE}$. This
                  suggest that there is a stable fixed point at the
                  intersection of the linear segments of the  excitatory and inhibitory nullclines.} \label{fig:13}
	\end{figure}

In addition to oscillatory activity in the middle range of rates, the
EI-population can exhibit non-oscillating asynchronous activity which
corresponds to a stable fixed point in the linear regime. Fig.\ref{fig:13} is the
simulation result of the population rates similar to the setup of the Fig.\ref{fig:11}
with higher $W_{II}$, which, as we will see later, makes the fixed point on
the linear section stable. 

%
%
%

\subsubsection{Logistic function approximation of gain functions}

In this section, we approximate gain functions by logistic functions to analyze bifurcation diagrams and approximate locations of bifurcaion points. For this purpose, we consider the  gain functions in following form:
\begin{align} 
g^x &(\rho _{Inh} , y_x) = \dfrac{\rho_{max}}{1 + \alpha(\rho_{Inh})e^{-ky_x}} - z_0, \nonumber \\
 y_x &= g_{syn}\tau w_{xI} \rho_{Inh} ( V_{Rinh} - V_{th})
  + g_{syn}\tau (w_{xE}\rho _{exc} + \rho_{Ext}^x) (V_{Rexc}-V_{th}) \nonumber \\& +g_{L}(V_{Le ak}-V_{th}), \nonumber \\
z_0 &= \dfrac{\rho_{max}}{1 + \alpha(0)e^{-kg_{L}(V_{Le ak}-V_{th})}} \label{eq:Log}
\end{align}
Here, $x$ stands for either excitatory (E) or inhibitory (I) gain functions, which have the same form but different input arguments. $\rho_{Ext}^x$ is the external excitatory input to the population $x$. 

\begin{figure}[h]
\centering
  \includegraphics[width=0.8\linewidth]{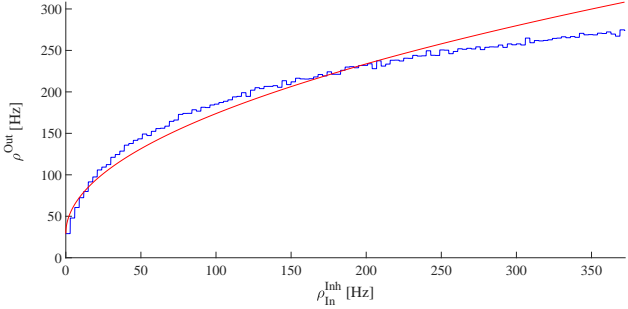}
  \caption[Output rate at threshold vs. input excitatory rate.]{Output rate as a function of input inhibitory rate, when the excitatory rate is selected in a way that average membrane potential of the neuron is $V_{th}$.  Neuron is operating near a saddle-node bifurcation point at which $F(I_{syn}) = k\sqrt{I_{syn}-I^*}$.} \label{fig:14}
\end{figure}

 At $y_x=0$, balanced input sets the membrane potential at the threshold value
 and  the output rate is approximately $g_{th} =
 \dfrac{\rho_{max}}{1+\alpha(\rho_{Inh})}$. Dependence of the output rate on
 inhibitory input, when the balance condition at threshhold holds, is
 represented by the function $\alpha$. At $y=0$, the output rate is
 proportional to the standard deviation in the input and it can be written as
 function of the inhibitory input rate as (see Fig.\ref{fig:14}): 
\begin{align}
   g_{th} = b_0 + b_1 \sqrt{\rho_{Inh}}
 \end{align}
  which fixes the function $\alpha(\rho_{Inh})$.
 
At equilibrium, the population rates satisfy:
 \begin{align} 
   & \rho_{I} =  g^I(\rho _{I} , c_{IE}\rho_{E} + c_{II}\rho_{I} + d\rho_{Ext}^I) - z_0 \nonumber \\
   & \rho_{E} =  g^E(\rho _{I} , c_{EE}\rho_{E} + c_{EI}\rho_{I} + d\rho_{Ext}^E) -z_0 \label{eq:RatesLog}
\end{align}
 where $c_{xy} = c k_{xy} w_{xy}(V_{R_y} - V_{th})$. As before, we take
 $w_{EE}$ and $\rho_{Ext}^E $ as control parameters. Therefore, the  solution
 of the first equation in Eq.\ref{eq:RatesLog} is independent of control parameters and gives a curve
 in the $\rho_{I}-\rho_{E}$ plane. Taking into account that the inverse of
 $g(\rho _{Inh} , y)$ is $g^{-1}(\rho _{Inh} , z) =\dfrac{1}{k}(
 \log (\dfrac{z}{\rho_{max} -z}) + \log (\alpha))$, the  equation for the inhibitory nullcline can be written as :
 
 \begin{align} 
   \rho_{E} = & \dfrac{1}{c_{IE}}(\dfrac{1}{k}[\log (\dfrac{(\rho_{I}+z_0)}{\rho_{max} -(\rho_{I}+z_0)}) + \log (\alpha)]  - c_{II}\rho_{I} - d\rho_{Ext}^E - g_L ) \label{eq:Inhnull}
 \end{align}
 
 The term in  brackets accounts for non-linearity in low and high values of
 $\rho_{I}$. The derivative of this term w.r.t.  $\rho_I$ is $
 \dfrac{\rho_{max}}{\rho_{I}(\rho_{max} -\rho_{I})}$, which is very small in
 the middle range of $\rho_{I}$ at values close to $0.5 \rho_{max}$. This is consistent with the fact that nullclines are approximately linear in the middle range of the rates.

\begin{figure}[h]
\centering
  \includegraphics[width=.9\linewidth]{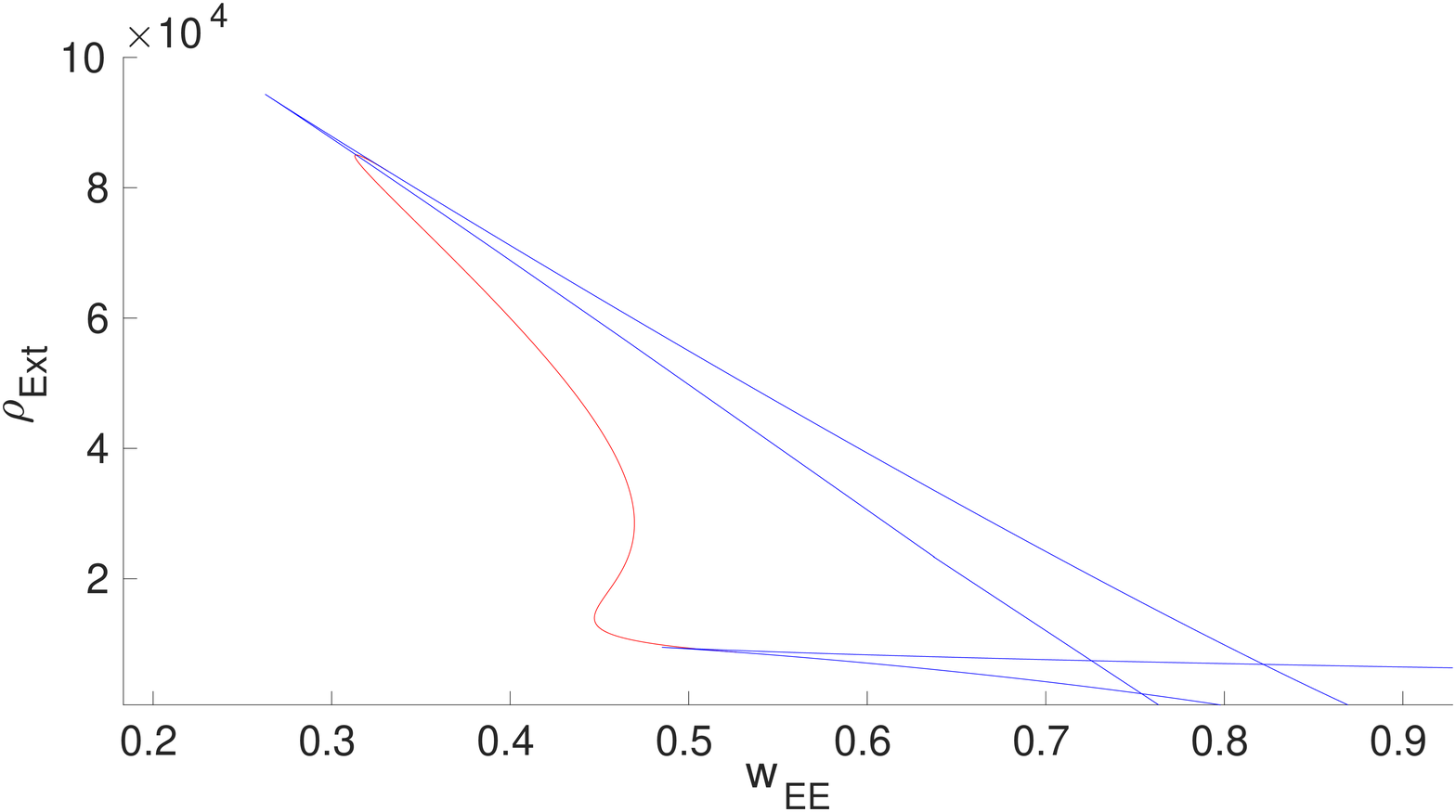}
  \caption[Local bifurcation diagram using Logistic gain function.]{Local
    bifurcation diagram in the control parameter plane $(W_{EE} ,
    \rho_{Ext})$. The red curve is the Hopf bifurcation line and the blue
    curves are saddle-node bifurcation lines. The free parameters of the model
    are $\rho_{Ext}^{Inh} =300Hz,W_{II}=1,W_{EI}=1.8$ and $W_{IE}=0.6$.} \label{fig:15}
\end{figure}

To analyze linear stability of the fixed points, we compute  derivatives
of the gain function:
\begin{align}
& \dfrac{\partial g^x}{\partial \rho_E} = kc_{xE} g^x(1-\dfrac{g^x}{\rho_{max}}) \nonumber \\
&\dfrac{\partial g^x}{\partial \rho_I} = kc_{xI} g^x(1-\dfrac{g^x}{\rho_{max}}) - \dfrac{1}{\alpha} g^x(1-\dfrac{g^x}{\rho_{max}})\dfrac{\partial \alpha}{\partial \rho_I}
\end{align}
Here $g^x$ stands for $g^I$ or $g^E$. One can substitute $\rho_I + z_0$ and $\rho_E + z_0$ from Equation \ref{eq:RatesLog} for $g^I$ and $g^E$, respectively.
Therefore, the Jacobian matrix components at the fixed point are:
  \begin{align} 
  J_{11} & = -1+c_{EE}\rho^E(1-\dfrac{\rho_E}{\rho_{max}}) \nonumber \\
  J_{12} &= c_{EI}\rho^E(1-\dfrac{\rho^E}{\rho_{max}}) - \dfrac{1}{\alpha}\rho^E(1-\dfrac{\rho^E}{\rho_{max}})\dfrac{\partial \alpha}{\partial \rho^I} \nonumber \\
  J_{21} &= c_{IE}\rho^I(1-\dfrac{\rho^I}{\rho_{max}}) \nonumber \\
  J_{22} &=-1+ c_{II}\rho^I(1-\dfrac{\rho^I}{\rho_{max}}) - \dfrac{1}{\alpha}\rho^I(1-\dfrac{\rho^I}{\rho_{max}})\dfrac{\partial \alpha}{\partial \rho^I}
 \end{align}
Hopf bifurcation occurs at fixed point solutions at which the  trace of the
Jacobian vanishes and its determinant is positive. On the other hand, at
saddle-node bifurcation occurs at points where the determinant vanishes. Figure \ref{fig:15} shows the arrangement of excitatory and inhibitory nullclines at Hopf and saddle-node bifurcation points. We proceed to approximate local bifurcation lines in the parameter space.    


The condition on zero trace $Tr(J) = J_{11}+J_{22}=0$ parameterized by the
inhibitory nullcline curve(Eq.\ref{eq:Inhnull}) determines the value for $w_{EE}$ at
which a Hopf bifurcation can occur. Ignoring terms related to $\alpha$, which are relatively small, from equating the trace to zero, we have

 \begin{align}
   & c_{EE}^H =  \dfrac{2 - c_{II} * \rho_I (1 - \dfrac{\rho_I}{\rho_{max}} )} { \rho_E (1 - \dfrac{\rho_E}{\rho_{max}}) }  \label{eq:HopfP}
 \end{align} 
 Then we have $w_{EE}^H = \dfrac{c_{EE}^H}{ck_{EE}(V_{RExc}-V_{th})}$. For each $(\rho_E ,\rho_I)$ point
 on the inhibitory nullcline (Eq.\ref{eq:Inhnull}), the above equation gives a value of
 $w_{EE}$ which sets the trace of the Jacobian to zero at this point. The
 second equation in Eq.\ref{eq:RatesLog} which corresponds to the excitatory nullcline
 determines $\rho_{Ext}^E $ parameterized by $\rho_{Inh}$. Next, we should
 check the  positivity of the determinant to sketch the Hopf bifurcation line
 in the $w_{EE}-\rho_{Ext}^E$ plane. Neglecting non-linearities caused by
 $\alpha$, the determinant of the Jacobian conditioned on zero trace is 
 \begin{align}
   det(J)_{\mid Tr(J)=0} = -(1 - c_{II} \rho_I (1 - \dfrac{\rho_I}{\rho_{max}}))^2 - c_{IE}c_{EI} \rho_I (1 - \dfrac{\rho_I}{\rho_{max}} )\rho_E (1 - \dfrac{\rho_E}{\rho_{max}})
 \end{align}
  At extremely low values of the rates (near zero), the determinant is
  negative because of the $-1$ in the above formula.  Conditioned on a
  sufficient amount of inhibitory feedback strength, which is proportional to
  $\lvert c_{IE}c_{EI} \rvert$, the determinant becomes positive at some point and the low
  fixed point loses stability through a Hopf bifurcation. A point where both
  determinant and trace of $J$ are zero, is called a \emph{Bogdanov-Takens} (BT) \emph{bifurcation point}.
 
  Inserting $\rho_E (1 - \dfrac{\rho_E}{\rho_{max}})$ from  $det(J)_{\mid Tr(J)=0}
  =0$ into the denominator of the formula for $w_{EE}^{H}$ and introducing
  the parameter $\gamma=\rho_I (1 - \dfrac{\rho_I}{\rho_{max}})$ , at the BT
  point in the low rate regime:
 \begin{align}
   & c_{EE}^{BT} =  \dfrac{\gamma c_{IE}c_{EI}(2- c_{II} * \gamma )} {(1-c_{II}\gamma)^2 } 
 \end{align}
 When $\mid c_{II}\gamma \mid$ is at a moderate value, i.e. sufficiently greater than one :
  \begin{align} 
 c_{EE}^{BT} \approx \dfrac{c_{IE}c_{EI}}{c_{II}} \label{eq:BTpoint}
 \end{align}
 
 If we take number of connections to satisfy  $\dfrac{k_{EI}}{k_{EE}} = \dfrac {k_{II}}{k_{IE}} $, then $w_{EE}^{BT} = \dfrac{w_{IE}w_{EI}}{w_{II}}$. On the semi-linear part of the inhibitory nullcline we have an  approximate
  relation between rates of the form  $\dfrac{\rho_I}{\rho_E} \approx
  \dfrac{w_{IE}}{w_{II}} $. The determinant on the line of zero trace when $\rho_E = \frac{w_{II}}{w_{IE}} := \beta \rho_I$ is 
  \begin{align}
   det(J)_{\mid Tr(J)=0} = -1 + 2c_{II} \gamma + (c_{ii}^2 - \beta c_{IE}c_{EI})\gamma^2 
 \end{align} 
The function $\gamma(\rho_I) $ has a maximum at $\dfrac{\rho_{max}}{2}$. With
this in mind, the condition for positive determinant at a potential Hopf
bifurcation fixed point at lower rates of the linear regime is
 \begin{align}
   det(J)^L_{\mid Tr(J)=0}  \approx -c_{II}^2 \rho_I^2 - c_{IE}c_{EI}\rho_E \rho_I \lvert c_{II} \rvert \rho_I^2( \lvert c_{EI} \rvert - \lvert c_{II} \rvert) > 0
 \end{align}
Therefore, if $\lvert c_{EI} \rvert> \lvert c_{II} \rvert$, the  Hopf bifurcation line survives in the
linear regime. On this line,  $w_{EE}^{Hopf} \approx \dfrac{2 -
  c_{II}\gamma}{c_{EE}^0\beta \gamma} $, which has negative derivative
$\dfrac{-2}{\beta \gamma^2}$. Thus, Hopf bifurcation in the linear regime
occurs at lower values of $w_{EE}$ compared to $w_{EE}^{BT}$ . On the other
hand, $\rho_{Ext}^E$ should increase to satisfy the fixed point condition of
equation \ref{eq:RatesLog} for the excitatory rate. Altogether, in the $(w_{EE} -
\rho_{Ext}^E)$ plane,  the Hopf bifurcation line extends to lower $w_{EE}$ and higher $\rho_{Ext}^E$ from the low BT to high BT.

 Ascending on the inhibitory nullcline, we reach the nonlinear high branch of
 the curve, where a linear relation between rates no longer holds and the
 second derivative of $\rho_E^*(\rho_I )$ will increase. In addition, $
 \gamma(\rho_{inh})$ decreases towards zero. Taking into account  these two
 facts, on the high branch $det(J)^L_{\mid Tr(J)=0}$ decreases and passes through
 zero at another Bogdanov Takens point at high rate values. If inhibitory
 feedback is not strong enough, the  conditions for Hopf bifurcation are not satisfied.
 
 To sketch the saddle-node bifurcation line we should look at solutions to
 $det(J) = 0$. Inserting $\rho_E(\rho_I)$ from Equation \ref{eq:Inhnull} into $det(J) $,
 for each point on the inhibitory nullcline, there exists some $w_{EE}$ for
 which  $det(J)=0 $. The only condition to check is $w_{EE}>0$. Again the
 condition that the  excitatory nullcline  intersects the inhibitory one at the fixed point determines $\rho_{Ext}$. 
 Along the semi-linear section of the nullcline, the condition $det(J)=0$
 translates into the alignment of the slopes of the linearized nullclines. Therefore, along this section $w_{EE}$ varies very little.

 Figure \ref{fig:15} shows Hopf and saddle-Node bifurcation lines with parameters written in the caption. As can be seen, there exist two Bogdanov -Takens bifurcation points at low and high values of external input corresponding to the intersection of nullclines in low and high firing rate regimes.

\begin{figure}[h]
\centering
  \includegraphics[width=1\linewidth]{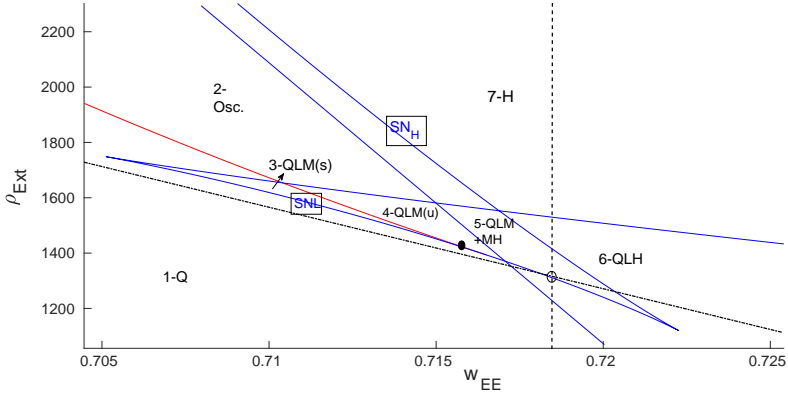}
  \caption[Zoom in on the local bifurcation diagram.]{Zoom in on the local
    bifurcation diagram at low firing rates and the corresponding regimes of
    phase space with different numbers of fixed points. The dashed line is the
    condition on the equal slope of linearized nullclines and the semi-dashed
    line is the condition on equal $y$-intercepts. The BT point (black dot) is
    close to the intersection of these lines. In the labeling of regions (Q)
    denotes the quiescent state fixed point, (L) is the fixed point at low
    firing rate, (M) is the fixed point in the linear section, and (H) is the high firing fixed point.} \label{fig:16}
\end{figure}

Figure \ref{fig:16} is the bifurcation diagram at low rates. Different regimes of phase
space corresponding to different numbers and/or types of fixed points have
been labeled. The system has between one and  five fixed points. Region (1)
with low values of $W_{EE}$ and external input strength is the quiescent state
with only one stable fixed point. In region (2), there is only an unstable
fixed point surrounded by a stable limit cycle corresponding to the
intersection of nullclines in the semi-linear sections. In regions (3) and (4)
near the BT point, two other fixed points exist at low firing rates. The type
of solution in these regions will be discussed later  in this section. Region (5) corresponds to the case where
there exist 5 intersection points on the nullcline map and the bi-stability of
the quiescent and the high state which survives after the annihilation of
unstable nodes on the middle section of the nullcline to the region (6). Finally, in the region (7), at high external input and synaptic weight, the only existing fixed point is the high firing one.  

  Dashed lines are the constraints of Equations \ref{eq:slope}-\ref{eq:yinter}
  corresponding to equal slope and $y$-intercept of the linearized
  nullclines. The vertical line is the value of $w_{EE}^*$ that matches the
  slopes, for $w_{EE}< w_{EE}^*$ the inhibitory feedback is getting
  stronger. The oblique line shows values of $\rho_{Ext}$ for each $w_{EE}$
  that equalize $y$-intercepts of linearized nullclines. In the region below this line $y_{inh}< y_{exc}$ and vice versa.

\subsubsection{Dynamics near the BT bifurcation point}

\begin{figure}[h]
\centering
  \includegraphics[width=0.7\linewidth]{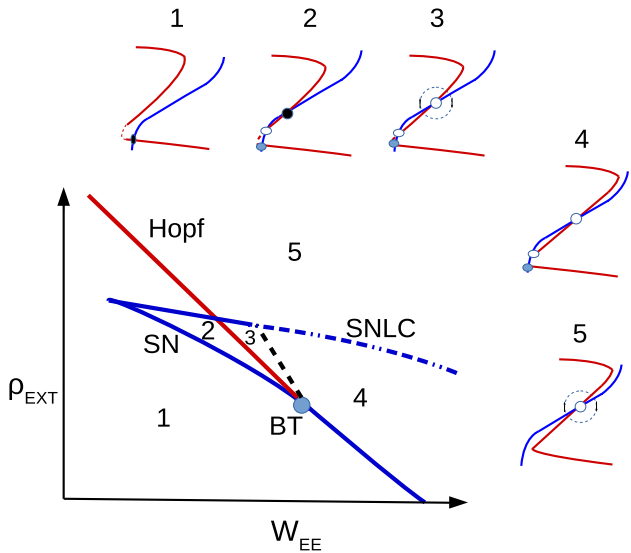}
  \caption[Nullcline arrangements near the BT point.]{Nullcline arrangements
    near the BT point. The black dashed line is the  saddle-separatrix loop
    bifurcation and the blue dotted-dashed is the saddle-node on limit cycle (SNLC) bifurcation line. } \label{fig:17}
\end{figure}

The exact locations of the $BT$ points $(c_{EE}^{BT},\rho_{Ext}^{BT} , \rho_{E}^{BT},\rho_I^{BT})$ are solutions of $det(J) =Tr(J) =0$ and $g_E(i_E)=g_I(i_I) = 0 $ . Figure \ref{fig:17} shows nullcline arrangements near the low BT point and the global saddle separatrix loop bifurcation line which annihilates the limit cycle solution of the region (3), shown in the same figure. 
In the previous section, we showed that the low BT point is located close to
the matching condition for the $y$-intercept and the slopes of the linearized
nullclines, which we rewrite here:
  \begin{align} 
  & c_{EE}^* = \dfrac{c_{IE}c_{EI}}{c_{II}} \nonumber \\
  & \rho_{Ext}^{E^*} = \dfrac{c_{EE}^*}{c_{IE}}(\rho_{Ext}^I-d) +d  \label{eq:BTlin}
 \end{align}
 where $d$ is a constant defined in Equation \ref{eq:yinter} . 
 
At the BT point the linearized matrix is of the form:
 
  \begin{align}
J_{BT} =  \begin{pmatrix} \alpha & -\beta \\ \dfrac{\alpha ^2}{\beta} & -\alpha \end{pmatrix}  \label{eq:JBT}
 \end{align}
 where $\dfrac{\beta}{\alpha} = \dfrac{c_{EI}}{c_{EE}} = \dfrac{c_{II}}{c_{IE}}$. At the  BT point, the Jacobian has a double zero eigenvalue and with proper coordinate transformation, it can be written in the form:
  \begin{align*}
  J = \begin{pmatrix}
0& 1\\
0 & 0
\end{pmatrix}
 \end{align*}
%
%
Consider the system  in the vicinity of the BT bifurcation point 
 \begin{align}
\dfrac{dz}{dt} = f(z,\mu) =Jz + F(z) , z , \mu \in R^2 \label{eq:DynSys}
	\end{align}	 
Suppose that at $\mu = 0$, the  system has a fixed point at $z_0$ with a Jacobian with a  zero eigenvalue of multiplicity two. At the BT point, there exist two generalized eigenvectors $q_0$ and $q_1$  such that
  	\begin{align*}
 Jq_0 =0 , Jq_1 = q_0
	\end{align*}
Also for $J^T$ we select vectors $p_{0,1}$ by 
\begin{align*}
 J^Tp_1 =0 , J^Tp_0 = p_1
	\end{align*}	 	
with normalization
\begin{align*}
 \langle p_0,q_0\rangle  = \langle p_1 ,q_1\rangle =1  \\ \langle
  p_0,q_1\rangle  = \langle p_1,q_0\rangle =0
	\end{align*}	 
	
 By linear change of coordinates with transformation matrix $T=(q_0,q_1)$ , i.e., $x = Tz$ , our system can be written as 
\begin{align}
\begin{pmatrix} \dot{x_1} \\ \dot{x_2}  \end{pmatrix} = \begin{pmatrix}
0 & 1 \\
0 & 0
\end{pmatrix} \begin{pmatrix} x_1 \\ x_2 \end{pmatrix} + \begin{pmatrix} f(x_1,x_2) \\ g(x_1,x_2) \end{pmatrix}
	\end{align}	 

Introducing a direction-preserving time reparametrization and smooth invertible parameter changes, we can tranforms the system to the normal form:
 \begin{align} 
 & \dfrac{y_1}{d\tau} = y_2  \nonumber \\
 & \dfrac{y_2}{d\tau}= \epsilon_1 + \epsilon_2 y_1 + a_2 y_1^2 + b_2 y_1y_2 + O(||y_1y_2||^3)  \label{eq:BTnormal}
	\end{align}	 	 
	$\epsilon_{1,2}(\mu)$ are transformed bifurcation parameters , $a_2 = g_{xx}/2$ and $b_2=g_{xy}+f_{xx}$ and $dt = (1+\Theta x_1)d\tau$.  The  $y$ coordinates relate to the original $z= T^{-1}x$ coordinates via : 
	\begin{align}
 &u_1 = x_1,  \quad  &y_1 = u_1 \nonumber \\
 &u_2 = \dot{x}_1,   \quad  &y_2 = u_2 + \Theta u_1 u_2  
	\end{align}	 	
where $\Theta= g_{yy}+2f_{xy} $. Fixed points of the normal form of Equation \ref{eq:BTnormal} are $(y_1 , y_2) = (\pm
\sqrt{(\dfrac{-\epsilon _1}{a_2}}, 0) $. Taking  $a_2 >0 $, when  $\epsilon_1 < 0$, there exist two fixed points, with Jacobian
\begin{align*}
 \begin{pmatrix}
0 & 1 \\
 \pm 2 \sqrt {\dfrac{-\epsilon _1}{a_2}} & \epsilon _2 \pm b_2  \sqrt  { \dfrac{-\epsilon_1}{a_2}}	\end{pmatrix}
	\end{align*}	 	

\begin{figure}
			\centering
			\includegraphics[width=1\linewidth,trim={0cm, 0cm, 0cm, 0cm},clip]{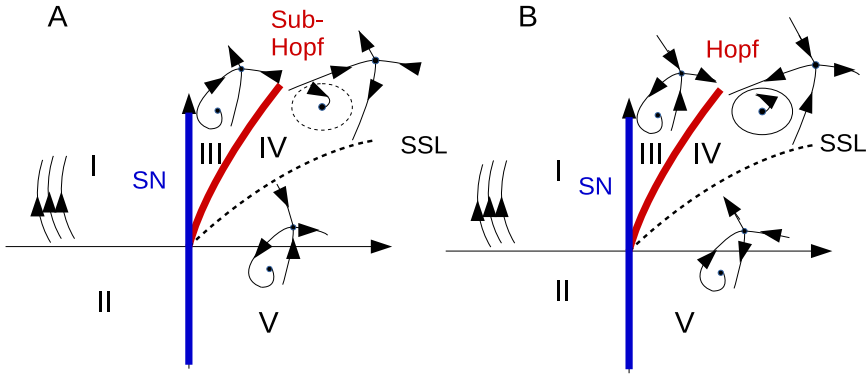}
		
		\caption[Phase space near BT points]{  Dynamic flow near
                  the high BT point (A) and   the low BT point (B). Blue lines are saddle-node bifurcations, red lines are Hopf bifurcations, and dashed lines are saddle-node separatrix loop bifurcations.} \label{fig:18}
	\end{figure}

 $y_0^+$ is a saddle in $\epsilon_1<0$ for all $\epsilon_2$,  while $y_0^-$ is
 a sink for $\epsilon _2< b_2 \sqrt{\dfrac{-\epsilon_1}{a_2}}$ and a source
 for $\epsilon_2 > b_2\sqrt  { \dfrac{-\epsilon_1}{a_2}}$. When $b_2 >0 $, 
 then the line $\epsilon_2 =  b_2 \sqrt{\dfrac{-\epsilon_1}{a_2}}$ is a
 sub-critical Hopf bifurcation and when $b_2 >0 $ the same line is a
 supercritical Hopf bifurcation. To summarize, defining  $\sigma =
 sgn(a_2*b_2)$, if $\sigma$ is negative then a stable limit cycle appears and
 the Hopf bifurcation is supercritical (Fig.\ref{fig:18}A), but if $\sigma $ is
 positive, we have a sub-critical Hopf bifurcation (Fig.\ref{fig:18}B). As shown in
 Figure \ref{fig:18}, near the BT point apart from local bifurcations, i.e., Hopf and
 saddle-node, there is saddle-node separatrix loop bifurcation which
 annihilates the  stable or unstable limit cycle that is produced by a super- or sub-critical Hopf bifurcation, respectively.

 Linearization near the BT point can help us to  identify regimes surrounding
 it without having to  calculate the  $\sigma$ parameter. Nullcline maps
 related to regions (2) and (3) in Fig.\ref{fig:17} shed light on the type of BT
 bifurcation.  In the plot corresponding region (3), $w_{EE}$ is higher which
 means that the Jacobian at the fixed point has lower determinant and higher
 trace. Of the two fixed points in regions (2) and (3) at the semi-linear
 section the one in the higher $W_{EE}$ regime is the unstable
 point. Therefore, in our case near the low BT point the phase space resembles
 the one in Fig.\ref{fig:18}B. Increasing $w_{EE}$ from region (2) will result in
 loss of stability of the fixed point in the linear branch by Hopf
 bifurcation, as the  trace of the Jacobian at the fixed point becomes
 zero. However, as we increase the $w_{EE}$, slope of the linearized
 approximation of the nullclines which are tangent to the stable and unstable
 manifolds of the saddle point that separate the quiescent fixed point and the
 limit cycle solution, get closer to each other. At some point, these
 manifolds cross over and therefore destroy the limit cycle solution through a
 saddle-node separatrix loop bifurcation and we end up with a fixed point of source type at the intersection of nullclines in the linear firing regime of region (4) in Fig.\ref{fig:17}.


	By writing the normal form we can analyze the type of BT from explicit linearization. For the case of $J_{BT}$ in equation \ref{eq:JBT}, generalized eigenvectors are  $q_0= \begin{pmatrix} 1 & \alpha/\beta \end{pmatrix}$ , $q_1 = \begin{pmatrix} 1 & (\alpha-1)/\beta \end{pmatrix}$ , $p_1 = (1 ,-\beta/\alpha)$ and $p_0= (1/\alpha -1 , \beta/\alpha)$. Therefore, new parameterized coordinates are  $x_1 =E/\beta - \alpha/\beta( E- I) $ and $x_2 =E - I$. The normal form parameters $a_2$ and $b_2$ are 
 \begin{align} 
& a_2 =  \dfrac{1}{2} < p_1 , F(q_0 , q_0)> \nonumber \\
& b_2 = <p_0 ,  F(q_0 , q_0)> + <p_1 , F(q_0 , q_1)>
	\end{align}	 	 
where $F^1(q_0,q_i) = \sum_{l,k} f_{lk} q_0^l q_i^k $ and $F^2(q_0,q_i) = \sum_{l,k} g_{lk} q_0^l q_i^k $.	Using the logistic gain function, second derivatives in $E-I$ coordinates can be written as
	\begin{align*}
& g_{EE} \propto W_{IE}^2 g_I(1-g_I)(1-2g_I)\\
& g_{EI} \propto -\lvert W_{II}\rvert \lvert W_{IE}\rvert g_I(1-g_I)(1-2g_I) \\
&f_{EE} \propto W_{EE}^2f_E(1-f_E)(1-2f_E) 
	\end{align*}	 	
where $g_{EI}< 0$, $f_{EE}>0 $  and $g_{EE}>0$. After straightforward but lengthy calculation we can confirm $sgn(\sigma) = sgn(a_2)sgn(b_2) <0$ using  $\lvert W_{II} \rvert g_I(1-g_I)=W_{EE}f_E(1-f_E)$  and $\dfrac{\beta}{\alpha} = \dfrac{c_{EI}}{c_{EE}} = \dfrac{c_{II}}{c_{IE}}$ at the lower BT point.
	
\subsubsection{Avalanches in the region close to the BT point } 	

We assume that the external input to both excitatory and inhibitory neurons is
dominated by the excitatory type and that  connections among excitatory
populations have a longer range. Therefore, the external excitatory input to
the excitatory population is higher than to the inhibitory one. On the other
hand, inhibitory connections are local and therefore, follow the dynamics of
the adjacent excitatory population. Strong local feedback provided by
inhibition prevents the excitatory network to be overloaded. However, it is
very closely balanced to set the network near the threshold of activation so
that the  system can respond efficiently to external input. In the
background  regime of spontaneous activity, the EI population  shows avalanche pattern dynamics and oscillatory behavior. Synchronization of oscillations and the scale-free avalanche dynamics are characteristic behaviors experimentally validated \cite{Beggs03, Gong, Meisel} . In the sequel, we will see that close to the BT at a low firing rate regime, we can observe both phenomena.  

 In the parameter space enclosed by Hopf and saddle-node bifurcation lines, i.e., region (4-QLM(u)) in Fig.\ref{fig:16}, there exist regions with both oscillatory and medium-range Poisson firing states. Decreasing $W_{EE}$ while changing $\rho_{Ext}^E $ accordingly, so that the low and medium fixed points move closer to the origin, the system moves towards the Bogdanov-Takens bifurcation point, where the saddle-node bifurcation and Hopf bifurcation lines intersect. In this regime, we see avalanche dynamics in our population.
 Close to the  BT point, the basin of attraction of the  quiescent fixed point
 shrinks and the noise level is high enough for escaping from it. This is in
 the adjacency of both the saddle-node bifurcation, which creates an unstable
 low and a weekly stable medium firing fixed point, and the Hopf bifurcation
 of the quiescent fixed point. This region corresponds to strong inhibitory
 feedback and sufficient imbalance in external excitatory input. In the
 nullcline graph, this translates into the state where the $y$-intercept of
 the excitatory graph is lower than the $y$-intercept of the inhibitory graph
 and the slope of the excitatory is larger than the slope of the inhibitory
 one. Increasing $W_{EE}$ causes the middle fixed point to move to higher
 rates and to have a larger basin of attraction. On the other hand, the saddle
 and the quiescent fixed point move towards each other in the phase diagram
 and annihilate each other at the saddle-node bifurcation.
 
\begin{figure}
			\centering
			\includegraphics[width=1\linewidth,trim={0cm, 0cm, 0cm, 0cm},clip]{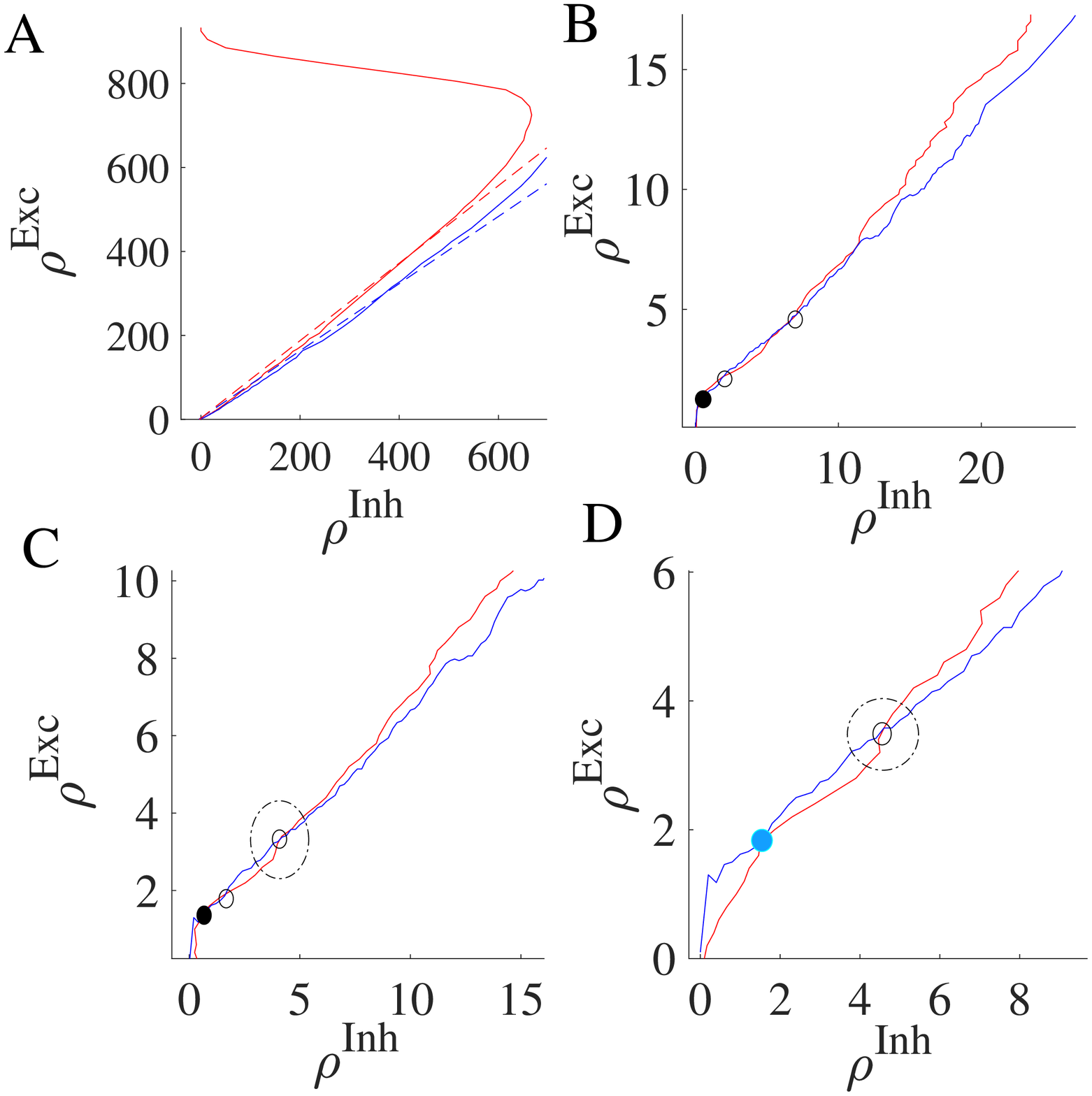}
		
		\caption[Nullclines' configurations around avalanche dynamic
                region.]{Nullcline configuration around the avalanche dynamic region. Red curves are excitatory nullclines and blue curves are inhibitory nullclines.} \label{fig:19}
	\end{figure}

Fig.\ref{fig:19} shows nullcline arrangements in the region where we observe avalanche
patterns. Fig.\ref{fig:19}A is the general position of nullclines
indicating the fixed point in the linear regime. The other three diagrams
correspond to two regimes near the BT point and transition between these
two. The diagram in Fig.\ref{fig:19}B  belongs to the section to the right of BT where
there exists a quiescent fixed point with a weakly unstable saddle in the
linear section. Here noise causes the system to escape from the basin of
attraction of the fixed point which then relaxes in the direction of the
nullclines. As nullclines lie on top of each other, the decay time is large
and the system shows high synchronous activity while returning to a quiescent
state. An increase of external drive or decrease of $W_{EE}$  leads to
saddle-node annihilation which leaves the system with a fixed point at the
middle section. Fig.\ref{fig:19}C  belongs to the state on the left side of
BT1 in the vicinity of Hopf bifurcation of the origin. In this case, there is
a limit cycle around the saddle point in the linear branch. Like the previous
case, adjacency of the fixed point at the  origin to the saddle shrinks the
basin of attraction of the quiescent state, and  therefore noise can bring the
system to the limit cycle which itself is sensitive to internal and external
noise.  Finally, Fig.\ref{fig:19}D  shows how saddle-nodes of the last two
diagrams are annihilated by saddle-node on limit cycle and saddle-node
bifurcations, respectively. Here a limit cycle solution emerges. However,
close to the  origin this limit cycle stays for a longer time in the lower section of very low firing because of slow flow in this region. The outcome is again a quasi-periodic burst of avalanches followed by a quiescent state.  

\begin{figure}
			\centering
			\includegraphics[width=1\linewidth,trim={0cm, 0cm, 0cm, 0cm},clip]{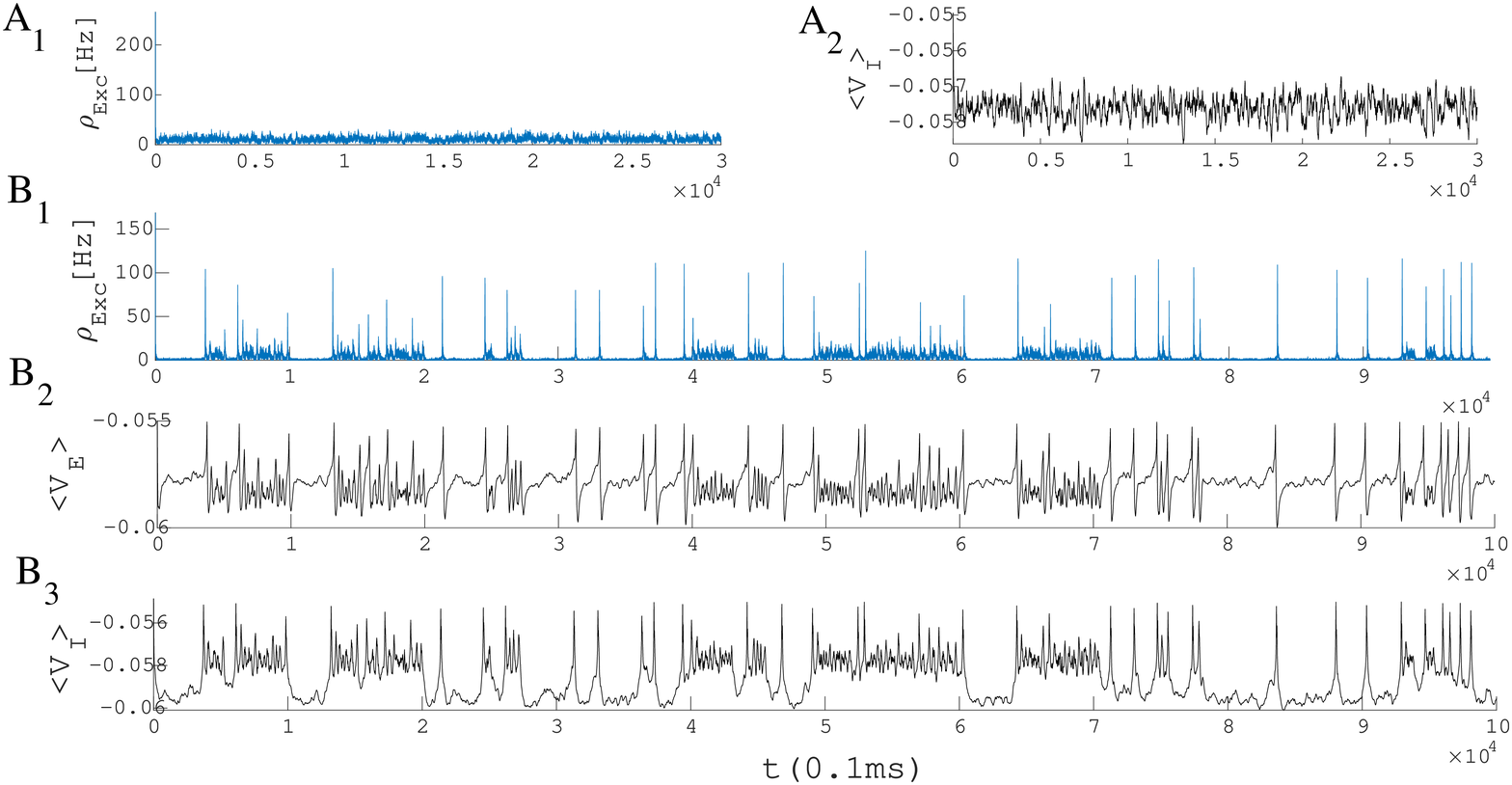}
		
		\caption[Avalanche style dynamics close to the BT point.(1)]{Avalanches  close to the BT point.
$W_{II} = W_{EI} = 2 , W_{IE}= 0.75$ , $\rho_{Ext}^{inh} = 150 Hz$ in both A and B plots. In (A) $W_{EE} = 0.65$ and $ \rho_{Ext}^{exc}= 218 Hz$ and in (B)  $ W_{EE} = 0.62 $ and $ \rho_{Ext}^{exc} = 223 Hz$. (B$_2$) Average membrane potential of excitatory population shows high fluctuation in the avalanches period. (B$_3$) Average membrane potential of inhibitory population shows high fluctuation in the avalanches period and two distinct level of polarization. In the quiescent state due to excess external current to the excitatory pool the average membrane potential of the excitatory population is slightly higher than the inhibitory one. }
		\label{fig:20}
	\end{figure}

\begin{figure}[h]
			\centering
			\includegraphics[width=1\linewidth,trim={0cm, 0cm, 0cm, 0cm},clip]{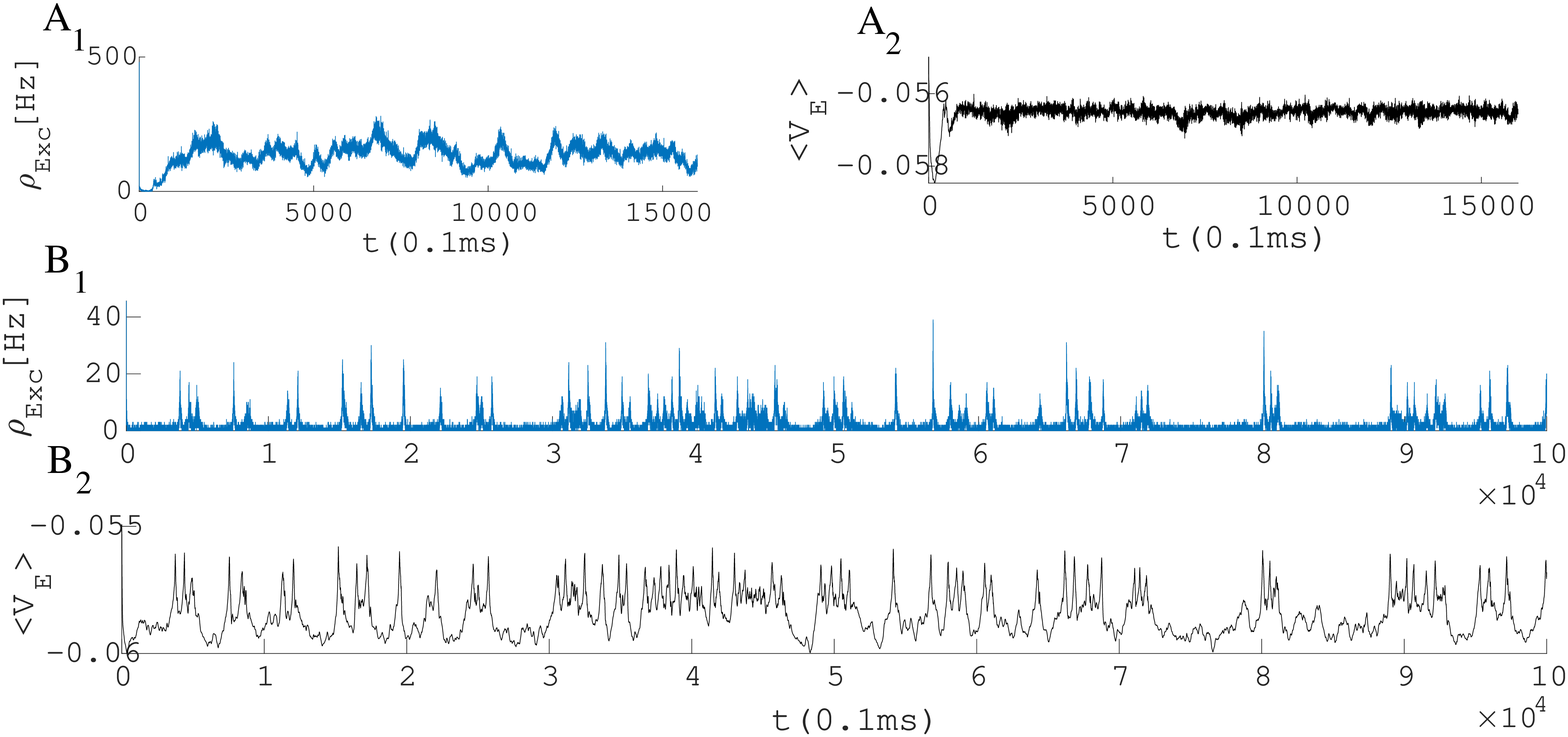}
		
		\caption[Avalanche style dynamics close to the BT point.(2)]{Same as Fig.\ref{fig:20} but with new parameters: $ W_{EI} = 1.5 ,W_{II}=2 , W_{IE}=0.75$  , $\rho_{Ext}^{inh}$= 150Hz , $\rho_{Ext}^{exc}$ = 230Hz  ,$W_{EE} = 0.55$ (A1-A2) and $W_{EE} = 0.52$ (B1-B2). 
} \label{fig:21}
	\end{figure}

 Fig.\ref{fig:20} shows avalanche characteristics of activity in parameter regime on the
 left of the BT point with the limit cycle solution very close to the origin
 ( region 3 in Fig.\ref{fig:17}). Finite size fluctuation leads to switch between these two states. In Fig.\ref{fig:20}A, $W_{EE}$ is higher and
 $\rho_{Ext}^E$ is slightly lower than the previous case and the system is
 located in the region with a fixed point in the low firing regime which is stable because of the high value of $W_{II}$ which corresponds to region (5) in Fig.\ref{fig:16}. 
Fig.\ref{fig:21}B shows avalanche dynamics on the right side of the BT point with an unstable fixed point in the linear section (region (4) in Fig.\ref{fig:17}). In both sets of figures increasing $W_{EE}$ moves the system out of the avalanche region with the difference that the fixed point at the linear section is stable in the first case and unstable in the second. Therefore,  the nearby regime of activity in the first case (\ref{fig:20}A) is a non-oscillatory inhomogeneous Poisson firing state while the corresponding regime near the second case is oscillatory (Fig.\ref{fig:21}A).

\subsubsection{Stability analysis of fixed points in the linear regime} 	

As we have seen in the last section, close to the BT point there exist regions
in which there is a low fixed point at the intersection of the semi-linear sections of nullclines.
The stability of the fixed point at the intersection of two nullclines is
determined by the Jacobian matrix of the linearized system, 
\begin{align}
A = \begin{pmatrix}
-1 + \dfrac{\partial f}{\partial E}  & -\lvert \dfrac{\partial f}{\partial I}\rvert   \\
\dfrac{\partial g}{\partial E}   & -1-\lvert \dfrac{\partial g}{\partial I} \rvert   
\end{pmatrix} 
\end{align}
Linear segments  intersect if  $y_{inh} < y_{exc}$ and $s_{exc} >s_{inh}$ or $y_{inh} > y_{exc}$ and $s_{exc} <s_{inh}$. When the slope and $y$-intercepts are equal, the Jacobian at the point of intersection is
\begin{align}
A = \begin{pmatrix}
a - \mu  & - (b -\dfrac{b}{a}\mu)  \\
a  & -b  
\end{pmatrix} 
\end{align}
with $\mu = a \dfrac{\rho_{Ext}^E-\rho_{Ext}^I}{d-\rho_{Ext}^I} $ . \\
 $a= \dfrac{\partial g}{\partial E}  = \alpha' W_{IE}K_{IE}  = \alpha W_{EE}k_{EE}-1+\mu$ 
 \\
 $d = \dfrac{g_{leak}(V_{rest} - V_{th}) }{\tau* g_{exc}^0*(V_{th}-V_{Rexc})}$ 
 \\
 $b= 1 + \lvert \dfrac{\partial g}{\partial I} \rvert  = 1 + \beta' W_{II}K_{II}  = (1-\dfrac{\mu}{a})^{-1}[\beta W_{EI}k_{EI}]$ 
 \\
 $\alpha = g_{exc}^0 \tau_{exc} \dfrac{(V_{th} - V_{Rexc})}{\sqrt{2\pi} \sigma_V^{Ex}}$
 \\
  $\alpha' = g_{exc}^0 \tau_{exc} \dfrac{(V_{th} - V_{Rexc})}{\sqrt{2\pi} \sigma_V^{Inh}}$
  \\
 $ \beta = g_{inh}^0 \tau_{inh} \dfrac{(V_{th} - V_{Rinh})}{\sqrt{2\pi} \sigma_V^{Exc}} $ 
 \\
 $ \beta' = g_{inh}^0 \tau_{inh} \dfrac{(V_{th} - V_{Rinh})}{\sqrt{2\pi} \sigma_V^{Inh}} $ 
 \\
 $\alpha * \beta' = \beta * \alpha '$
 \\

Because external excitatory input to the excitatory population is greater than
to the inhibitory population and inhibitory connections are assumed to be local, $\mu$ is slightly positive. Define $E= \rho_E- \rho_E^p$ and $I = \rho_I - \rho_I^p $, where $\rho_I^p$ and $\rho_E^p$ is the fixed point location at the linear poisson regime with $\rho_I^p \approx \dfrac{b}{a} \rho_E^p$. 

At $ \mu = 0 $ the eigenvalues of $A$ are $0$  and  $a-b$ with corresponding
eigenvectors  $u_1 =(\dfrac{b}{a} ,1) $ and $u_2 =(1,1) $. By coordinate
transformation to $u_1 $ and $u_2$ coordinates, we can write down the dynamics
in the decoupled system as
\begin{align}
 \dot u =  \begin{pmatrix}
0  & 0  \\
0  & a-b  
\end{pmatrix} u
\end{align}
where,
 \begin{align}
 u =  \begin{pmatrix}
\dfrac{b}{a}  & 1  \\
1  & 1  
\end{pmatrix}^{-1} \begin{bmatrix}
           E  \\
           I  \\
         \end{bmatrix} = \dfrac{a}{a-b} \begin{bmatrix}
           I-E  \\
           E- \dfrac{b}{a}I  \\
         \end{bmatrix}
\end{align}

with the transformed initial condition 
\begin{align*}
 u_0 = \dfrac{a}{a-b} \begin{bmatrix}
           I_0 -E_0  \\
           E_0 - \dfrac{b}{a}I_0  \\
         \end{bmatrix}
\end{align*}
which has the following solution in $u$ coordinates
\begin{equation} 
 u(t) = \dfrac{a}{a-b} \begin{bmatrix}
           I_0 -E_0  \\
          ( E_0 - \dfrac{b}{a}I_0) e^{(a-b)t}  \\
         \end{bmatrix}
\end{equation}

Back into $(E,I)$ coordinates:
\begin{align} 
 \begin{bmatrix}
           E(t)  \\
           I(t)  \\
         \end{bmatrix} = \dfrac{a}{a-b} ( I_0 -E_0)  \begin{bmatrix}
           \dfrac{b}{a}  \\
           1  \\
         \end{bmatrix}  +   \dfrac{a}{a-b}( E_0 - \dfrac{b}{a}I_0) e^{(a-b)t} \begin{bmatrix}
           1 \\
          1  \\
         \end{bmatrix} \label{eq:flowBT}
\end{align}

\begin{figure}
			\centering
			\includegraphics[width=0.5\linewidth,trim={0cm, 0cm, 0cm, 0cm},clip]{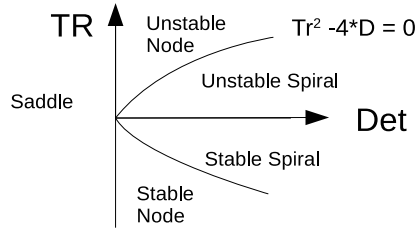}
		
		\caption[Stability of fixed points in the linear regime.]{Stability of fixed points in the linear regime based on values of trace and determinant of the Jacobian.} \label{fig:22}
	\end{figure}

So for this linear system, when $a-b <0$, the  initial imbalance of excitatory and
inhibitory input leads to a stationary relation of the form $E = \dfrac{b}{a} I $. 
Now, consider the case in which the linearized nullcline slopes are slightly different with the  Jacobian
\begin{align} 
A = \begin{pmatrix}
a - \mu  & - (b +\epsilon)  \\
a  & -b  
\end{pmatrix} 
\end{align}
Here $TR= \lambda_1 + \lambda_2 = (a-b)-\mu $ and $det= \lambda_1 \lambda_2= a\epsilon + \mu b$. Based on the sign of determinant and trace of the Jacobian at the fixed point, stability is determined (Fig.\ref{fig:22}).
Under the condition that $b + \mu >a  $ and $\epsilon > -\dfrac{b}{a}\mu$, both eigenvalues are negative : $\lambda_1 = \dfrac{b\mu - a\epsilon}{b-a} $  and  $\lambda_2 = (a-b) + \dfrac{2a(\epsilon -\mu)}{a-b} $.  We also have $\lvert \lambda_1 \rvert<< \lvert \lambda_2 \rvert$ for small differences in the slopes.
Eigenvectors corresponding to these eigenvalues are 
\begin{align*}
& u_1 = ( \dfrac{b}{a} +\lambda_1 , 1) \\
& u_2 = (1 + \lambda_2 , 1)
\end{align*}
Therefore, the dynamics in the linear regime can be projected to the slow stable manifold $u_1$. One can approximately write down the evolution of the rates as in Equation \ref{eq:flowBT}.

 $\epsilon > -\dfrac{b}{a}\mu$  corresponds to the case that the slope of the
 excitatory nullcline is higher than of the inhibitory nullcline (stronger
 inhibitory feedback $W_{EI}W_{IE} > W_{II}W_{EE}$ )  and the $y$-intercept of
 the excitatory nullcline is slightly lower, i.e.  stronger external excitatory input to the excitatory population than to the inhibitory one. Moreover, this is the case when
 $W_{II}$ is high enough to guarantee the  $b>a$ condition. When all these
 requirements are met, the fixed point in the linear segment is stable and we
 observe an asynchronous low to medium firing state as in Fig.\ref{fig:13} and
 Fig.\ref{fig:20}. Around this regime, an increase in $W_{EE}$ will increase $\mu$ and a
 change in $\rho_{Ext}^E$ moves the fixed point along the linear section. The
 intersection in the linear regime transcends to higher rates by increasing
 $W_{EE}$. This lets the  determinant  decrease while the trace increases, which eventually  destabilizes the fixed point.
In the vicinity of the low BT point, based on the value of $W_{II}$, in the linear section either a weakly stable or a weakly unstable fixed point surrounded by a limit cycle appears. In both cases, the eigenvalue close to zero with eigenvector $u1$ governs the slow dynamics around these points. 

 Consider the case of imaginary eigenvalues of the Jacobian, $\lambda_{\pm} = \sigma \pm i\omega	$ with eigenvectors $v_{\pm} = v_r \pm v_i$, which satisfy
	  	\begin{align*}
&	A [ v_r v_i] = [v_r v_i] \begin{pmatrix}
\sigma & \omega \\
-\omega & \sigma
\end{pmatrix}
	\end{align*}
	By defining the tranformation matrix $T = [ v_r v_i]$, the linearized matrix is $ Q = T^{-1}AT = \begin{pmatrix}
\sigma & \omega \\
-\omega & \sigma
\end{pmatrix} $ and the solution of the linear system is of the form
\begin{align*}
e^{At} x_0 = T e^{\sigma t}\begin{pmatrix}
cos(\omega t) & sin(\omega t) \\
- sin(\omega t) & cos(\omega t)
\end{pmatrix} T^{-1} x_0
	\end{align*}
 By using the coordinate transformation $ u = T^{-1}x $, we can write the
 evolution  $\dot{u} = Qu$ with $u_0 = T^{-1}x_0$. The linearized dynamic
 predicts damped oscillations of frequency $\omega=\sqrt{det -
   \dfrac{Tr^2}{4}}$ when $\sigma<0$ and at the  Hopf bifurcation point when
 $\sigma=0$ the  frequency of oscillations will be $\omega=\sqrt{det_{H}}$. At
 the nullcline intersections of linear segments close to the Hopf bifurcation,
 the oscillation frequency is close to the imaginary part of the eigenvalues: $\sqrt{det - \dfrac{Tr^2}{4}}$.

Along the slow manifold, the inhibitory and excitatory rates vary linearly as
$I = \dfrac{a}{b} E \approx  \dfrac{k_{ee}W_{ee}}{k_{ei}W_{ei}} E$. This
relation balances the average current for each population. Therefore, near the
BT bifurcation point, the dynamic of slow field, $E-I$, can be written as 
	\begin{align} 
	\dfrac{d(E-I)}{dt} =  \epsilon(E-I) + c (1-\dfrac{a}{b})^{-1}(E-I)^2 \dfrac{1}{\sqrt{N}}(1-\dfrac{a}{b})^{\dfrac{1}{2}}(E+I)^{\dfrac{1}{2}} \eta(t) \label{eq:slowfield}
	\end{align}
where $\epsilon $ is close to zero, the first nonlinear term of the Taylor expansion has been taken into account and $\eta(t)$ is a white noise added to the microscopic equation based on the Poisson firing assumption. 


\subsubsection{Characteristics of avalanches}

\begin{figure}
			\centering
			\includegraphics[width=1\linewidth,trim={0cm, 0cm, 0cm, 0cm},clip]{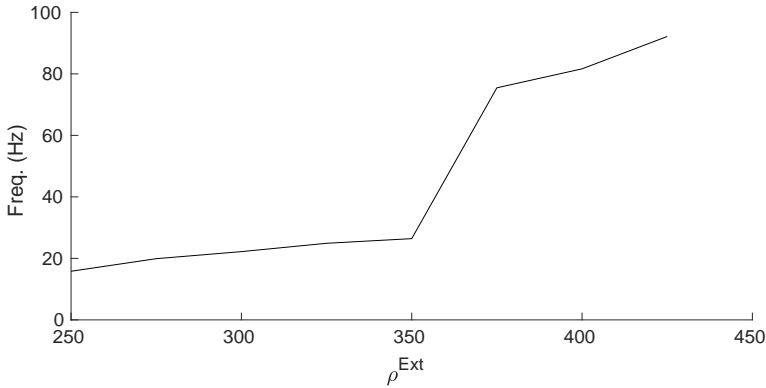}
		
		\caption[Frequency of avalanches.]{Frequency of avalanches and
                  oscillatory activity increase by input strength. In lower
                  values of excitatory external input, the limit cycle
                  solution is very close to the origin and the system shows
                  avalanches. By increasing the  external drive, the limit cycle moves away from the origin (quiescent state) and becomes stable. Oscillations have a higher frequency at higher external input rates with semi-linear relations in both regions. 
} \label{fig:23}
	\end{figure}

 For the values of $W_{EE}$ near the BT point at the low firing rate, there
 exists a range of external input strength for which the firing pattern is
 quasi-periodic with excitatory avalanches followed by inhibitory ones. The
 mean escape time from the basin of attraction of the quiescent fixed point
 reduces when the external input increases, and thus, the frequency of
 avalanches increases. Further increase of external input leads to stability
 loss of the quiescent state and appearance of higher frequency oscillations
 in the medium range of rates (see Fig.\ref{fig:23}).

\begin{figure}
			\centering
			\includegraphics[width=1\linewidth,trim={0cm, 0cm, 0cm, 0cm},clip]{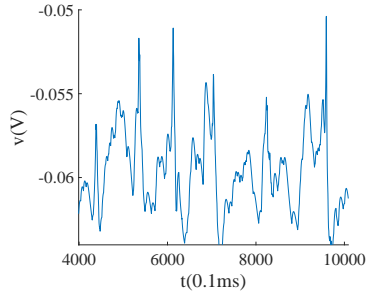}
		
		\caption[ Membrane potential track of a single neuron during avalanche dynamics.]{ Membrane potential track of a single neuron during avalanche dynamic of Fig.\ref{fig:20}. Avalanches at population level can be seen as periods of rising potential in the individual neurons which are sustained longer than avalanche range due to slow synaptic decay. Individual neurons do not fire in every single avalanche.  } \label{fig:24}
	\end{figure}

   In the avalanche regime,  the membrane potential shows sub-threshold
   oscillations as can be seen in Fig.\ref{fig:20} and Fig.\ref{fig:21}. In the down phase of the
   cycle, neurons stay near the resting potential while at the up-state they reside closer to the threshold, but at a distance  that permits high variability of firing. The membrane potential of a single neuron is depicted in Fig.\ref{fig:24}, which shows aperiodic firing and up-down states of membrane potential.

\begin{figure}
			\centering
			\includegraphics[width=1\linewidth,trim={0cm, 0cm, 0cm, 0cm},clip]{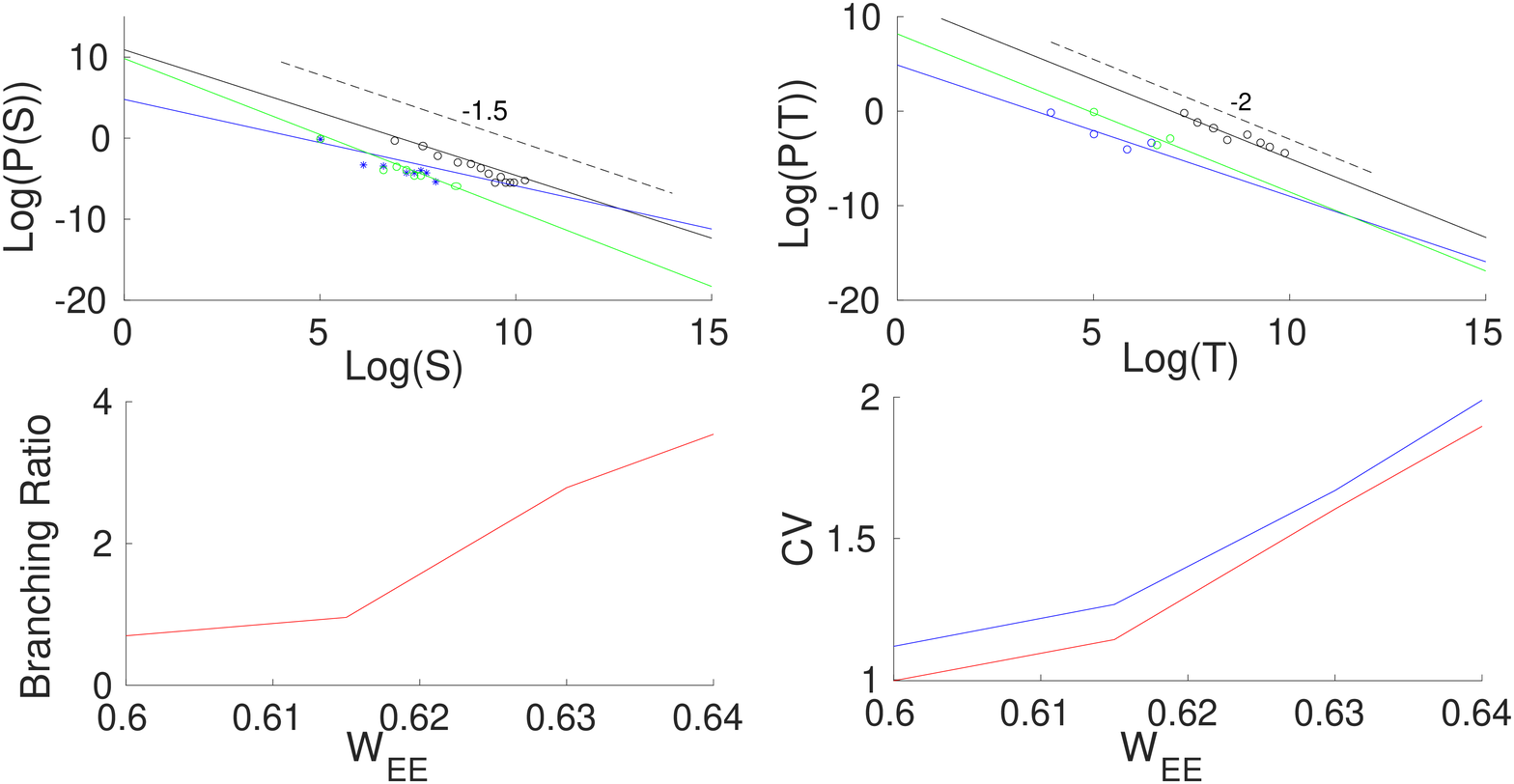}
		
		\caption[Avalanche characteristics(1).]{Same as Fig.\ref{fig:20} with $W_{EI} = 2 ,W_{II}=2 , W_{IE}=0.75$  , $\rho_{Ext}^{inh}$= 150Hz , $\rho_{Ext}^{exc}$ = 230Hz  with different values of $W_{EE} \in (0.6,0.65)$.  Avalanche sizes (A) and duration(B) distributions in log-log plot with linear fit. Green line ($W_{EE}$ = 0.63), black ($W_{EE}$= 0.615) and  blue ($W_{EE}$=0.6). (C) Branching ratio and (D) CV of firing time intervals of individual neurons (red for the excitatory neurons and blue for the inhibitory ones)}\label{fig:25}

	\end{figure}

\begin{figure}
			\centering
			\includegraphics[width=1\linewidth,trim={0cm, 0cm, 0cm, 0cm},clip]{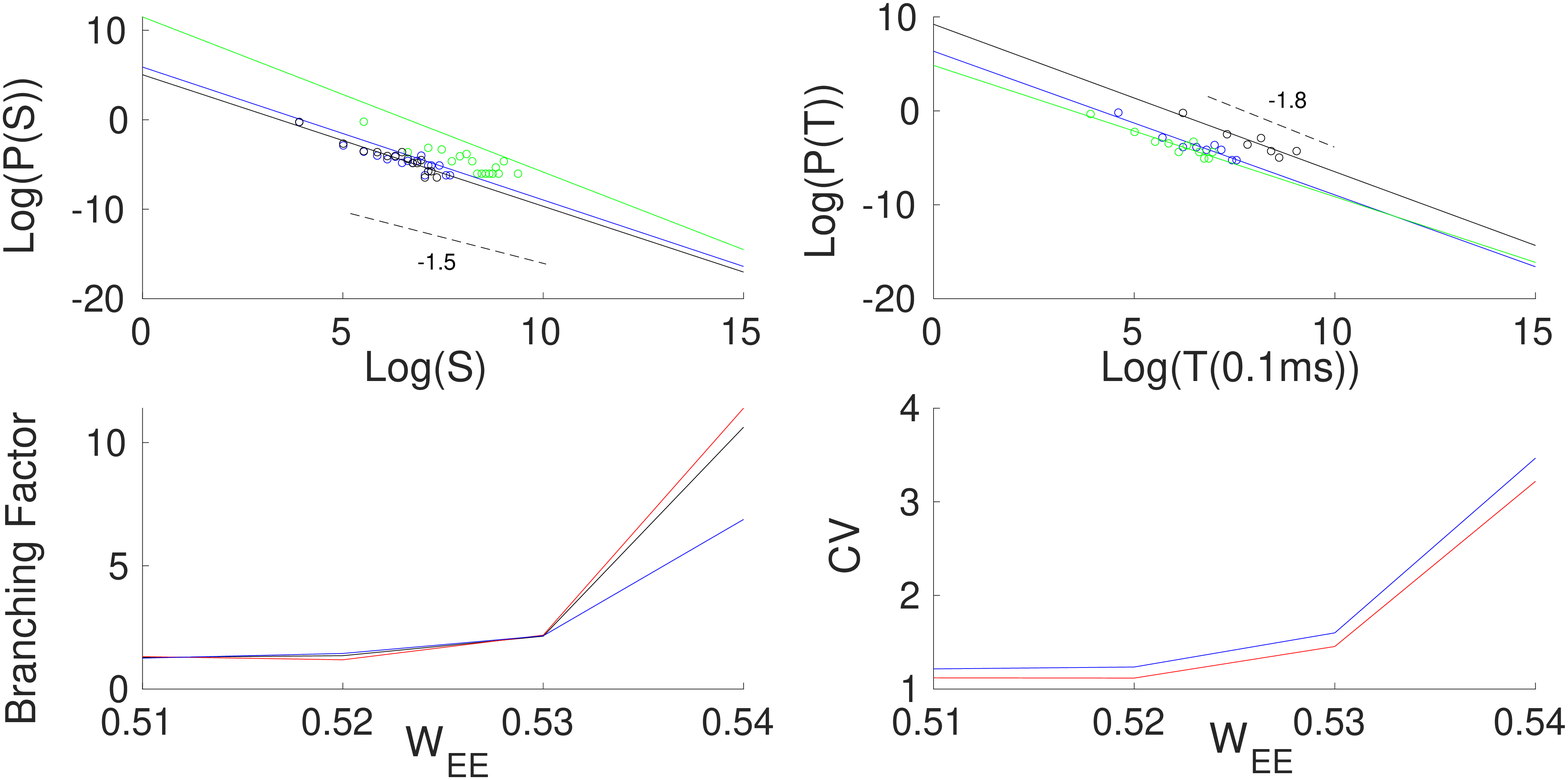}
		
		\caption[Avalanche characteristics(2).]{Characteristics of avalanches for the model with $W_{EI} = 1.5 ,W_{II}=2 , W_{IE}=0.75$  , $\rho_{Ext}^{inh}$= 150Hz , $\rho_{Ext}^{exc}$ = 2300Hz and $W_{EE} \in (0.516,0.54)$  as in Fig.\ref{fig:21} . (A-B) Avalanche sizes and duration distributions in log-log plot with their corresponding least square error linear fit. Green line ($W_{EE}$ = 0.54), black ($W_{EE}$=0.53) and blue ($W_{EE}$ = 0.51). (C) Branching ratio and (D) CV of interspike time intervals of individual neurons (red for the excitatory neurons and blue for the inhibitory ones). \label{fig:26} }

	\end{figure}

While avalanches occur quasiperiodically, in most of them only a fraction of
neurons fire.  As shown in (Fig.\ref{fig:25}D and Fig.\ref{fig:26}D) neurons fire with CV close to
one in the lower $W_{EE}$ regime, close to the BT point.
Variability in the size of avalanches is another interesting item to
investigate. The size distribution of avalanches has a longer tail approaching
the BT point. It follows a power-law distribution for   avalanche size  $P(S)
\propto S^{-\tau}$ with slope $\tau = -1.5$  close to the BT point, see Fig.\ref{fig:25}A
and Fig.Fig.\ref{fig:26}A. Further away from the critical point, avalanches have
characteristic average size and their size probability density moves away from
the power-law distribution. Furthermore, the probability distribution for the
duration of avalanches follows a power law with an exponent close to $\eta = -2$ near the BT point(see Fig.\ref{fig:25}B and Fig.\ref{fig:26}B).

The branching ratio can be defined as the average number of postsynaptic
neurons of a specific neuron that fire by receiving the synaptic current from
that neuron. The branching ratio can be an indicator of scale-free avalanche
dynamics. When inhibition and excitation are balanced and the system resides
near a quiescent state, the branching parameter stays close to, but below one which is an indicator of stronger inhibitory feedback. As can be seen in Fig.26 and Fig.27, this value is lower in the paramter regime close to the BT point and becomes $>1$ further away from it.

 Here, we assume that by synchronous activation of $n_E$  neurons the postsynaptic neurons which are connected to these neurons will receive both excitatory and inhibitory currents caused by the synchronous input. Each neuron receives a fraction $k_{EE}$ of excitatory and $k_{EI}$ of inhibitory currents produced by active neurons. The average potential change among neurons will be
\begin{align}
\langle \Delta V \rangle = & \langle k_{EE} n_E\rangle \dfrac{1}{C} g_0 w_{EE}  \tau  (V_{Rexc}- V_{E} )  + \langle k_{EI} n_I \rangle \dfrac{1}{C} g_0 w_E  \tau  (V_{Rexc}- V_{E} ) 
	\end{align}	
  Close to the bifurcation point, there exists a tight dynamic balance between
  excitatory and inhibitory rates, following Equation \ref{eq:flowBT}, which sets
  $\langle \Delta V \rangle =0$. Based on the assumption that neurons fire
  with Poisson statistics, we can write the variance of the  potential change in the postsynaptic neuron pool as
 
 \begin{align} 
 \langle \Delta V ^2\rangle =  \tau^2 g_0^2 ( \langle k_{EE}n_E\rangle w_{EE}^2   (v_{Rexc} - V_{th})^2 + \langle k_{EI}n_I\rangle (w_{EI})^2( v_{Rinh} - V_{th})^2 )   \label{eq:varAv}
	\end{align}	

On the other hand, the number of postsynaptic neurons that fire by receiving
an increase in voltage of value $\Delta V$ is
\begin{align} 
\sigma =  N_{Exc} \int_{V_{th} - \Delta V}^{V_{th}} P(V,t=\infty)\approx & - \dfrac{N_{exc}\Delta V^2}{2}\dfrac {\partial p(v^E ,t=\infty)}{\partial v} \mid _{v^E = V_{th}} \label{eq:sigma}
	\end{align}	

From equation \ref{eq:app1-0} in Appendix \ref{app:1}, for the stationary probability density we have 

\begin{align} 
&\dfrac {\partial p(v^E ,t=\infty)}{\partial v}\mid_{v^E = V_{th}} = - \dfrac{2 C^2\rho_{exc}}{ D_e( v_{Rexc} - V_{th})^2   +  D_i( v_{Rinh} - V_{th})^2} \label{eq:parv}
	\end{align}

Inserting Equation \ref{eq:parv} in Equation \ref{eq:sigma} and averaging $\sigma$ over different realizations of the synchronous firing using Equation \ref{eq:varAv} and dividing by $\langle n_E \rangle$ leads to 
\small
 \begin{align} 
\sigma^E   \approx  \dfrac{  \tau^2 g_0^2 [w_{EE}^2 \rho_{exc}^{st}  (v_{Rexc} - V_{th})^2 + \rho_{exc}^{st} \dfrac{\langle n_I\rangle}{ \langle n_E \rangle} w_{EI}^2( v_{Rinh} - V_{th})^2 ]} { D_e( v_{Rexc} - V_{th})^2   +  D_i( v_{Rinh} - V_{th})^2 } \label{eq:BR}
	\end{align}	
\normalsize
The average number of active inhibitory and excitatory neurons $\langle n_I\rangle $ and $\langle n_E \rangle$,  relates to stationary rates as $\dfrac{\langle n_I \rangle}{\langle n_E \rangle} = \dfrac{ \rho_I}{\rho_E}$. Inserting this relation into equation \ref{eq:BR}, we find out that the branching ratio is close to one near the BT point. Because of slightly stronger inhibitory feedback, it is slightly below one.

Excitatory neurons stay in a low firing regime with average membrane potential
close to the middle point between the firing threshold and the
resting-state potential, i.e., at $V \sim -57mv$. At this point, a sufficient
fraction of neurons is close to the threshold, whose activation can cause a
series of firing. On the other hand, inhibitory neurons, which have a lower
stationary membrane potential because of lower external input, provide
negative feedback with a delay that depends on the resting initial state and
the strength of the connection between inhibitory and excitatory
sub-networks. The dynamic balance of excitation and inhibition in the linear
UP state leads to  critical behavior. As average currents to the cells are
balanced far from the firing threshold, fluctuations in these currents have a
larger effect and therefore, the size of events and their durations ar emore variable.

\begin{figure}[h]
\centering
  \includegraphics[width=1\linewidth]{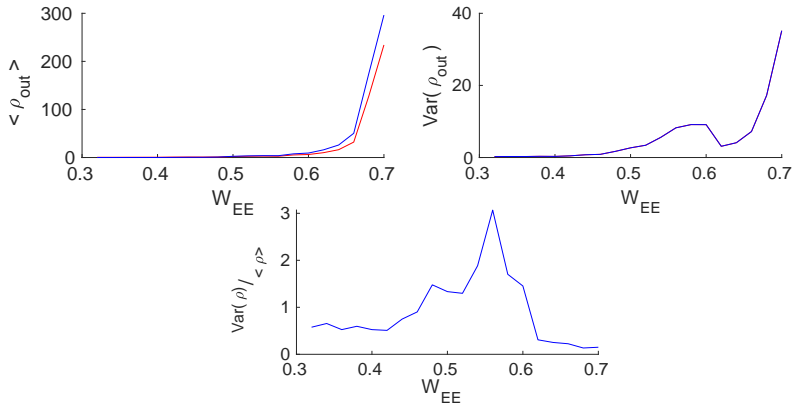}
  \caption[Phase transition in EI population.]{Stationary population rates (A)
    (Blue for Inh. and Red for Exc.), variance (B) and normalized variance (C) for EI
    population vs. $W_{EE}$ at the fixed value of $\rho_{Ext} =250Hz$. Other
    parameters were set to $W_{IE}= 0.75$, $\rho_{Inh}=150Hz$, $W_{II}=W_{EI}=2$.} \label{fig:27}
\end{figure}

\begin{figure}[h]
\centering
  \includegraphics[width=1\linewidth]{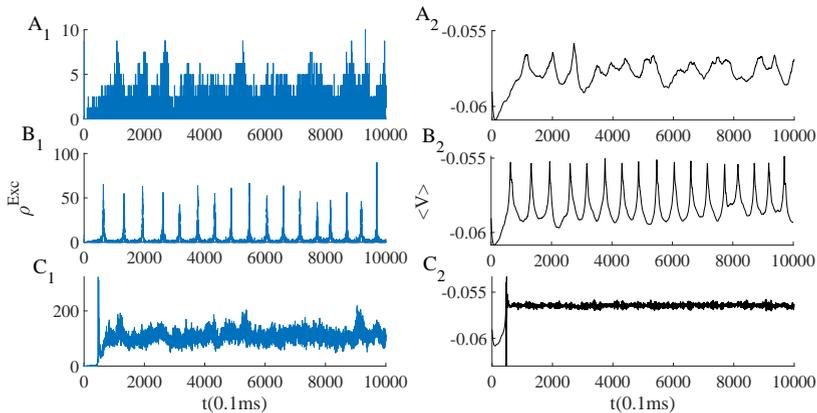}
  \caption[Phase transition in EI population(2).]{Excitatory population rate
    (Right plots) and average membrane potential (Left plots) when the system
    is  slightly below the  Hopf bifurcation point (A), slightly above Hopf
    bifurcation point (B), or after the saddle-node bifurcation point (C) corresponding to the states in Fig.\ref{fig:27} with the same parameters.} \label{fig:28}
\end{figure}

Moreover, let us consider the onset of avalanche dynamics in the EI population
receiving external input with fixed rates by selecting $W_{EE}$ as the only
dynamic parameter (see Fig.\ref{fig:27}). By increasing $W_{EE}$, a second-order phase
transition happens at the Hopf bifurcation. Around this value, the normalized
variance of the population rate is maximized and oscillations appear in the
system. In Fig.\ref{fig:28}, this happens at the value $W_{EE} \approx 0.57$. Further
increase of $W_{EE}$ results in the saddle-node bifurcation which produces a stable high firing rate state at values around $W_{EE} \approx 0.67$.
	
Although the activity is noise-driven, the state of the system depends on synaptic weights, which determine the response to the external input. There must be a self-organizing mechanism, which in a wider range of  input strengths and initial configurations of synaptic weights tunes the system close to the BT point.

%
%
%
%


\section{Discussion}
We have seen that in a large sparse network of spiking neurons the input to
the cells in the state of asynchronous firing is Poisson and investigated
conditions on Poisson firing at the single neuron level. We chose the
conductance-based leaky integrate and fire model to take the strong dependence
of the inhibitory postsynaptic current on the voltage level into consideration. Next, we introduced linearization of the neuron gain function in the Poisson firing regime and presented a linear Poisson neuron model which we used to analyze interconnected networks of  excitatory and the inhibitory neurons.

 The network of spiking neurons with the assumptions of homogeneity, large
 size, and sparse connectivity can be modeled by the dynamics of the mean
 fields. The excitatory and inhibitory mean-field equations are a set of
 nonlinear equations with free parameters including  the average synaptic strength between different types of neurons. Taking a set of these free parameters as control parameters of the model one can analyze the bifurcation patterns in the system. Here, we chose the excitatory external drive and the synaptic weight from excitatory to excitatory neurons as control parameters. The latter regulates the strength of the inhibitory feedback in the local population and the former controls the level of forced activity from other populations. The qualitative picture of the bifurcation patterns does not change by the choice of different synaptic weights as control parameter. 
In analyzing the bifurcation diagram, we are mainly interested in the loss of
stability of the quiescent state. This can happen through a saddle-node or  a
Hopf bifurcation by either increasing the external drive or $W_{EE}$. At a
certain point called the Bogdanov-Takens point, the saddle-node and Hopf
bifurcation lines meet. Near this point there is a tight balance of the
inhibitory and the excitatory average currents to the cells. This balance
cancels out much of the high amplitude excitatory and inhibitory currents to
each cell  and causes the average membrane potential of the neurons in the
population to stay away from the threshold. In this regime, the activity of
the spiking neurons is fluctuation driven which makes the firing time
intervals highly variable.  In this case, the statistics of the firing is
close to a Poisson point process matching the experimental findings. On the
other hand, the balance of excitation and inhibition leads to avalanche style
dynamics near the BT point. Slow oscillations emerge at the Hopf bifurcation
line and through a saddle-node bifurcation, a pair of low firing stable and unstable fixed points comes into existence.

The next step after identifying the operating dynamical regime that produces
the desired output is to investigate mechanisms that can tune the parameters
of the system at the desired region of the phase space. The interplay of
dynamics and structural organization on different spatiotemporal scales is
coordinated and framed by multi-scale self-organization mechanisms that
emerged in the organism during evolution. There exist opposing forces that
shape the structure and activity such as excitation and inhibition currents
produced by excitatory and inhibitory neuron types, depression and
potentiation of the connections between neurons, modulation of the
concentration of chemicals, and homeostatic considerations on energy
consumption and information processing performance. Balancing and coordinating
these opposing forces might be one explanation for the scale-free
characteristic of spontaneous activity. We shall investigate the self-organization by spike-timing dependent plasticity and short term synaptic depression in another article.

\bmhead{Acknowledgments}
ME wants to thank International Max Planck Reseach School for funding his position during the research period.

\section*{Declarations}

\begin{itemize}
\item Funding: Open access funding provided by Max Planck
Society.
\item Conflict of interest: The authors declare that the research was conducted in the absence of any commercial or financial relationships that could be construed as a potential conflict of interest.

\item Availability of data and materials :The original contributions presented in the study are included in the article/supplementary material, further inquiries can be directed to the corresponding author.
\item Code availability :The code used in this work is available for download through github under the following link:
\item Authors' contributions:ME and JJ designed research. ME performed
research. M.E. wrote the manuscript. J.J. edited the
manuscript. All authors reviewed the manuscript, contributed to the article, and approved the submitted version.
\end{itemize}

\bibliography{sn-article}

\newpage
\begin{appendices}

\subsection{Gaussian approximation for response to the Poisson input } \label{app:1}

The Fokker-Planck equation corresponding to Eq.\ref{eq:Gauss} in section \ref{sec3-sub1} in the It\^o interpretation is

\begin{align} \label{eq:A1.1} 
\dfrac {\partial p(v,t)}{\partial t} = & - \dfrac{1}{C} \dfrac{\partial}{\partial v} [ (a - bv) p(v,t)] + \dfrac{1}{C^2}\dfrac {D_e}{2}\dfrac{\partial ^2}{\partial v^2} ( v_{Rexc} - v(t))^2  p(v,t) \nonumber \\ & +
 \dfrac{1}{C^2}\dfrac {D_i}{2}\dfrac{\partial ^2}{\partial v^2} ( v_{Rinh} - v(t))^2  p(v,t) \equiv  - \dfrac{\partial }{\partial v} J(v,t) \tag{A1.1}
	\end{align}
with the boundary condition  $p(V_{th} ,t) = 0$. The density current at $v =
V_{th}$ is equivalent to the firing rate, and this current is fed back to the
equation at $v =V_{r}$ resulting in a discontinuity of the  membrane potential derivative.  
	\begin{align*}
	 J(v_{r}^+ , t) - J( v_{r}^- , t ) = J(v_{th} , t-t_{ref} ) \equiv r(t-t_{ref})
	\end{align*}
The stationary probability distribution and firing rate are obtained by solving the equation
\begin{align*}
J_{st}(v) =  r_0 \Theta ( v - V_{r})
	\end{align*}	
which results in 
\begin{align} \label{eq:app1-0} 
 & \dfrac{1}{C} \{a - bv + \dfrac{1}{C} D_e ( v_{Rexc} - v) + \dfrac{1}{C} D_i ( v_{Rinh} - v) \} p(v) \nonumber \\
 & -\dfrac{1}{C^2}\{ \dfrac{D_e}{2}( v_{Rexc} - v)^2  +  \dfrac {D_i}{2}( v_{Rinh} - v)^2 )\} \dfrac{d p(v)}{d v}  =r  \tag{A1.2}
	\end{align}
for $V_{r}<v < V_{th}$. Together with the normalization requirement $ \int
_{-\infty}^{V_{th}} p(v) dv + r_0*t_{ref} =1 $ one can solve equation \ref{eq:app1-0}.
numerically for both the stationary probability distribution and the stationary firing rate.  

\begin{figure}
   \specialcaption{$A1.1$}{}
			\centering
			\includegraphics[width=1\linewidth,trim={0cm, 0cm, 0cm, 0cm},clip]{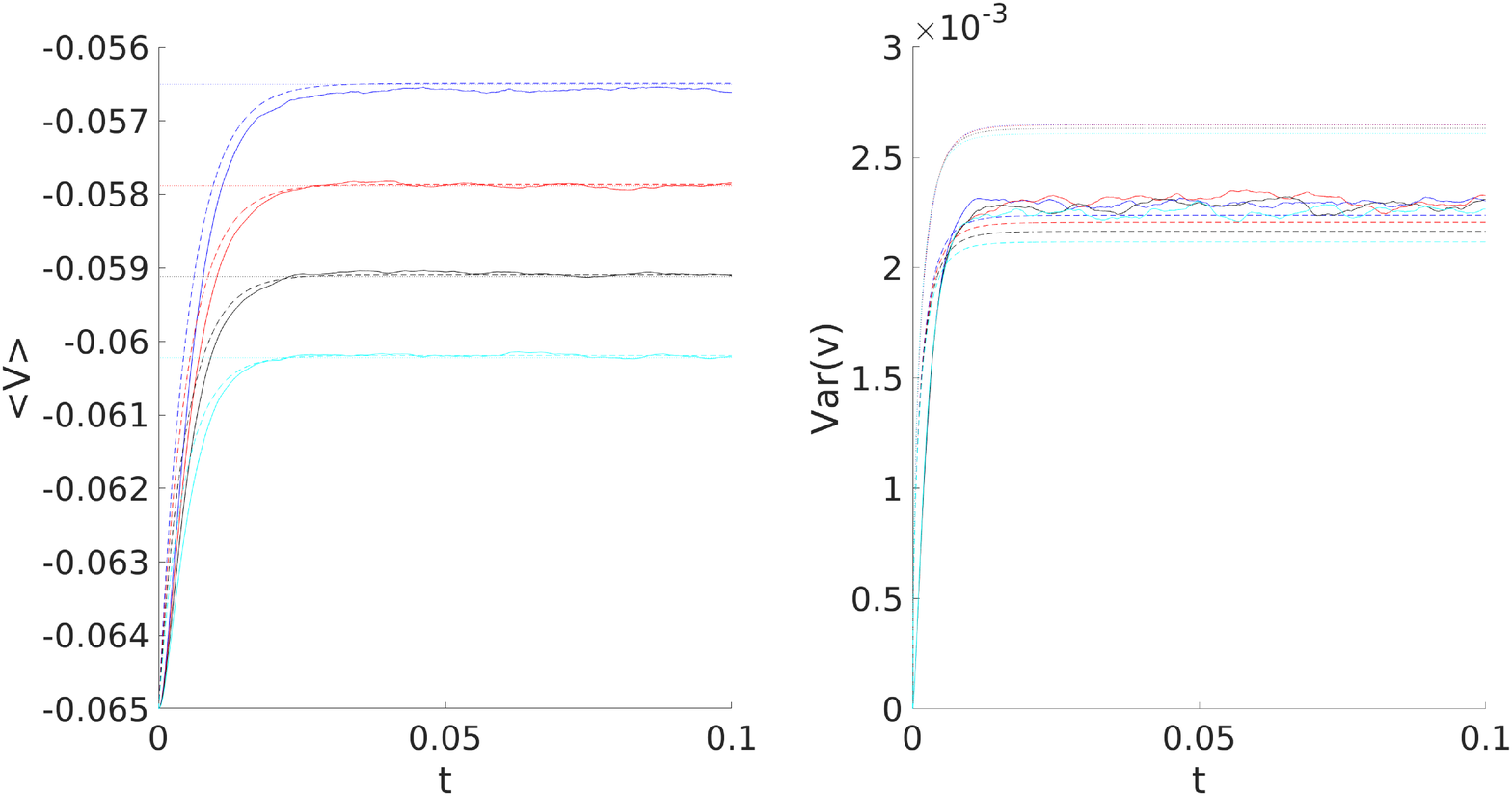}
		\caption[Membrane potential variance and mean evolution using $\tau$-expansion.]{Evolution of the mean and the variance of the potential distribution of a population of neurons each receiving a fixed equal inhibitory input rate but different excitatory rates. Dashed lines are the trajectories determined from Gaussian noise approximation and the dotted lines are derived from the tau expansion method. The approximation by tau expansion improves the estimation of variance.($\tau_{syn} =2.5 ms$) } \label{fig:A1.1}
		 
	\end{figure}

When $V_{st}$ is sufficiently smaller than $V_{th}$ , i.e., in the low firing rate regime, we can ignore the non-linearity caused by the threshold and write down the evolution of the mean and variance of the membrane potential as follows (See Fig.\ref{fig:A1.1})

\begin{align*}\
\dfrac {d \langle v(t)\rangle}{dt} =&  \dfrac{1}{C} (a - b\langle v(t)\rangle )\\
\dfrac {d Var(v,t) }{dt} = & -\dfrac{2b}{C} Var(v,t) + \dfrac{1}{C^2}[ D_e \langle (V_{Rexc} - v(t))^2 \rangle + D_i \langle (v(t)-V_{Rinh})^2 \rangle ] \\
 =&( -\dfrac{2b}{C} +  \dfrac{D_e + D_i}{C^2})Var(v,t)  + \dfrac{1}{C^2}[ D_e (V_{Rexc} - \langle v \rangle )^2 + D_i (V_{Rinh} -\langle v \rangle)^2]
	\end{align*}

This leads to the stationary value for average and variance of the membrane voltage 

 \begin{align}\label{eq:A1.3}
 \langle v \rangle_{st} = & \dfrac{a}{b} \nonumber \\
 Var(v)_{st} =  & \dfrac{1}{2bC-(D_e +D_i)}[ D_e (V_{Rexc} -\langle v\rangle_{st})^2 + D_i (\langle v \rangle _{st} - V_{Rinh})^2] \tag{A1.3}
\end{align}

\begin{figure}
		 \specialcaption{$A1.2$}{}
			\centering
			\includegraphics[width=1\linewidth,trim={0cm, 0cm, 0cm, 0cm},clip]{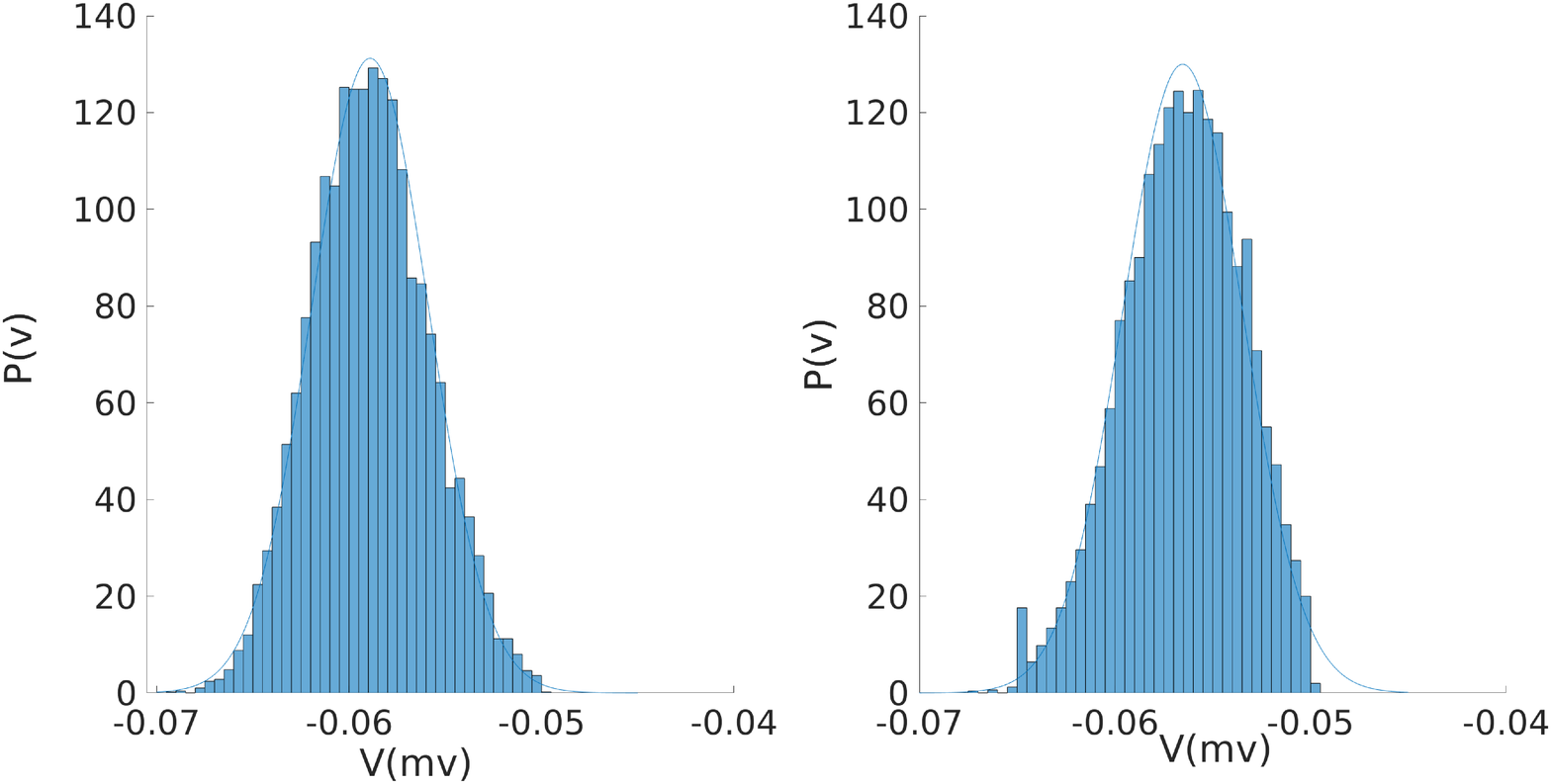}
		
		\caption[Gaussian approximation of membrane potential probability density] {Membrane potential distribution and Gaussian approximation (Blue lines) for two different sets of inhibitory and excitatory Poisson inputs. The average membrane potential values are $-0.59mv$(Left) and $-0.56mv$(Right)} \label{fig:A1.2}
	\end{figure}

In the low firing regime, the stationary probability distribution can be approximated as (see Fig.\ref{fig:A1.2}): 
 \begin{align*}
 P_{st}(V) = & \dfrac{1}{\sqrt{2\pi}\sigma_{V(st)} }exp( -\dfrac{(V -\langle V\rangle)^2}{2 \sigma^2}) \\&+ c \delta (V - V_{Rest})      \qquad V<V_{th}  \\
 P_{st}(V) = & 0  \hspace{1.2 in} V \geq  V_{th} 
\end{align*}

The stationary firing rate is derived from equation  \ref{eq:app1-0}  by plugging in the
Gaussian approximation for the stationary potential probability density  $P(V,t \to \infty) = N( \langle V\rangle , \sigma _{V(st)}) $:

 \begin{align} \label{eq:App1}
 & r = \dfrac{1}{C^2} \dfrac{1}{2} D(V_{th})^2 \dfrac{d p(V)}{d V}_{\mid V = V_{th}} \nonumber \\ &=  -\dfrac{b}{C} \dfrac{D(V_{th})^2}{D(\langle V \rangle)^2} * \dfrac{(V_{th} - \langle V \rangle)}{\sigma _{V(st)} \sqrt{2\pi}} exp( -\dfrac{(V_{th}-\langle V \rangle)^2}{2 \sigma _{V(st)}^2)}) \nonumber \\
 & \approx \dfrac{b}{\sqrt{\pi} C}  \dfrac{(V_{th} - \langle V \rangle)}{\sqrt{2}\sigma _{V(st)} } ( 1-  \dfrac{(V_{th} - \langle V \rangle)^2}{2 \sigma _{V(st)^2}}) \tag{A1.4}
\end{align}

\subsection{Tau expansion} \label{app:2}

The equation $\dfrac {dv(t)}{dt} = f(v) + \eta (t) g(v) $ where $\eta (t) $ is colored Gaussian noise with correlation 

\begin{align*}
 \langle \eta (t) \eta (t') \rangle = \dfrac{H}{\tau} e^ {\dfrac{-(t-t')}{\tau}  }
\end{align*}	
corresponds to the following Fokker-Planck equation derived by expansion with respect to $\tau$ :
\begin{align*}\
\dfrac {\partial p(v,t)}{\partial t} =  & -\dfrac{\partial}{\partial v} [ f(v) p(v,t)] +  H \dfrac{\partial}{\partial v} g(v) \dfrac{\partial}{\partial v} \{ g(v)[1 + \tau  g(v) (\dfrac {f(v)}{g(v)})'] p(v,t) \} \\
 =  &-\dfrac{\partial}{\partial v} [ (f(v)+ Hg(v)g'(v) ) p(v,t) ]+   H \dfrac{\partial ^2}{\partial v^2} \{g(v)^2p(v,t)\}  \\
 & -\tau H \dfrac{\partial}{\partial v}\{ g'(v)(g(v)f'(v) - g'(v)f(v))P(v,t) \}  \\
 &+\tau  H \dfrac{\partial^2}{\partial v^2} \{ g(v)(g(v)f'(v) - g'(v)f(v)) p(v,t) \}
	\end{align*}	

In our case $H_e = \dfrac{\tau^2g_0^2w_e^2\lambda_e}{2} = \dfrac{D_e}{2}$, resulting in:
 
\begin{align} \label{eq:A2.1}
\dfrac {\partial p(v,t)}{\partial t} = & - \dfrac{1}{C} \dfrac{\partial}{\partial v}\{ ( a - bv +F(v))p(v,t) \} \nonumber \\
& +  \dfrac{D_e}{2C^2} \dfrac{\partial ^2}{\partial v^2} \{G_e(v)p(v,t)\} +  \dfrac{D_i}{2C^2}\dfrac{\partial ^2}{\partial v^2} \{G_i(v)p(v,t)\} \tag{A2.1}
\end{align}	
where

\begin{align*}
F(v) = &\dfrac{D_e}{2C} [-2(v_{Rexc} - v(t)) - \dfrac{\tau}{C} (a - bV_{Rexc}) ] \\ & + \dfrac{D_i}{2C} [-2( v_{Rinh} - v(t)) - \dfrac{\tau}{C} (a - bV_{Rinh}) ] \\	
G_e(v)= & (v_{Rexc} - v(t))^2 +  \dfrac{\tau}{C}  ( v_{Rexc} - v(t))(a - bV_{Rexc}) \\
G_i(v) = &  (v_{Rinh} - v(t))^2 + \dfrac{\tau}{C} ( v_{Rinh} - v(t))(a - bV_{Rinh})	
\end{align*}	

\subsection{Neuron response in higher values of input rates and synaptic time constant} \label{app:3}
As it can be seen from Fig.\ref{fig:A1.1} and Fig.\ref{fig:3}C in section \ref{sec3-sub1}, the stationary standard
deviation of the membrane potential in the case of Poisson input does not show
high sensitivity to input rates when the stationary mean potential is fairly
away from the threshold. In Fig.\ref{fig:3}, increasing the excitatory input rate
by $60\%$ causes a $2\%$ increase of $\sigma(V)_{st}$ on average for different values of $\langle v \rangle _{st}$.
This can be predicted from equation \ref{eq:A1.3} as both the input noise and drift
term in the numerator and the denominator depend linearly on input
rates. Suppose $\langle v \rangle _{st}$ is fixed for a set of inhibitory and
excitatory rates, which means there is a linear relation of the form $\rho_E =
\kappa \rho_I +c$ originating from the condition on the fixed stationary
average membrane potential. In the limit of high rates, the stationary variance approaches a constant value:
 \begin{align} \label{eq:app3-1}
  & Var(v)_{st} =  \dfrac{\sigma ^2}{2bC} = \dfrac{ \alpha - \dfrac{\beta}{\lambda _E}}{ \gamma - \dfrac{\eta}{\lambda _E}} \nonumber \\
  \dfrac{\alpha}{\gamma}= &\dfrac{ \tau g_0 (V_{Rexc}-\langle V_{st}\rangle)(\langle V_{st} \rangle- V_{Rinh})}{V_{Rexc}-V_{Rinh}} * [w_E (V_{Rexc}-\langle V_{st} \rangle)+ w_I (\langle V_{st} \rangle- V_{Rinh})]  \nonumber \\ 
 \dfrac{\beta}{\eta} = &\dfrac{w_i \tau g_0 (\langle V_{st}\rangle- V_{Leak})(\langle V_{st}\rangle -V_{Rinh})^2}{  \langle V_{st}\rangle - V_{leak}} = w_i\tau g_0 (\langle V_{st}\rangle -V_{Rinh})^2  \tag{A3.1}
\end{align}
when $\dfrac{\alpha}{\gamma} > \dfrac{\beta}{\eta}$, the variance of the
stationary membrane potential (and hence also the output rate) increases by proportional increase of both
inhibitory and excitatory rates and reaches a constant value
$\dfrac{\alpha}{\gamma}$.  This condition translates into $ w_E
(V_{Rexc}-\langle V_{st}\rangle)^2 > w_I (V_{Rinh}-\langle
V_{st}\rangle)^2$. As long as $ \langle V_{st}\rangle $ is adequately lower
than $V_{th}$, $ w_E (V_{Rexc}-\langle V_{st}\rangle) \approx w_I
(V_{Rinh}-\langle V_{st}\rangle)$ and the condition mentioned above holds
because $( V_{Rexc}-\langle V_{st}\rangle)$ is greater than
$(V_{Rinh}-\langle V_{st}\rangle)$ by a factor of about  $2$. 

We remind the reader  that the Gaussian approximation is only legitimate in
the case of small $\tau$  and for a low firing rate regime. We want to
consider cases in which these two conditions are not satisfied. Firstly, at
higher values of $\tau_{syn}$ and at the stationary values of the average
membrane potential lower than the threshold (low firing regime), there is an
inversion of the mentioned scenario. In this case, at sufficiently high values
of the input rates, conditioned on constant average membrane potential, the
variance of the membrane potential and accordingly the output rate decrease
(Fig.\ref{fig:3}A). This is due to the filtering effect of the input by the gradual
decay of the synaptic conductances. Using the
method of $\tau$-expansion of Appendix$\ref{app:2}$, we will consider autocorrelation in synaptic
conductance and partially take into account this effect. The filtering of
the high-frequency signal by the slow conductances is a mechanism of gain
control. In general, the decay and the rise time of the inhibitory synapses
are longer than the excitatory ones which would highlight the inhibitory input
as its overall strength  increases through temporal persistence. In addition, the voltage-dependent inhibitory current is higher at higher values of the membrane potential. The balanced average membrane potential is somewhere in the mid-range. Longer synaptic decay time constant, higher synaptic strength, delay, and potential dependence of the inhibitory synapses increase the overall inhibition strength and compensate for the smaller number of them in comparison with excitatory synapses.  
It should be noted that output rates on the constant voltage level line with
balanced input at $V=-57mv$ vary linearly with the input rates at moderate
rate values corresponding to the low firing rate regime (Fig.\ref{fig:3}A).

On the other hand, when the stationary average membrane potential is located
right at the threshold value, in conflict with the low firing regime
assumption, equation \ref{eq:app3-1} does not hold and higher rates of balanced input
lead to a higher output rate independent of the value of $\tau_{syn}$
(Fig.\ref{fig:3}B). Moreover, the output rate varies like $\sqrt{I_{in}}$. For analyzing the low firing rate, we could linearize the output rate around the midpoint of the neuron potential range, i.e., at $V=-57mv$. 

\subsection{ First passage time  } \label{app:4}

To determine first passage time(FPT) statistics let us first review the
recursive formula for FPT moments as discussed in Siegert
\cite{Siegert51}. Suppose the stochastic process $X_t$ with conditional probability density $P(x,t \mid x_0)$ satisfies the Fokker-Planck equation 
\begin{align*}
\dfrac {\partial p(x,t \mid x_0)}{\partial t} =  - \dfrac{\partial}{\partial x} [ A(x) p(x,t \mid x_0)]  +  \dfrac{1}{2}\dfrac{\partial ^2}{\partial x^2} (B(x) p(x,t \mid x_0))
	\end{align*}
with initial and boundary conditions $ P(x,t_0 \mid x_0) = \delta(x-x_0)$ and $
P(\pm \infty,t \mid x_0) = 0$, respectively. The first passage time probability density $\rho(\Theta \mid t , x_0)$  of the  stochastic process following the above Fokker-Planck equation  satisfies
\begin{align*}
\rho(\Theta \mid t , s)  = -2 \dfrac{\partial}{\partial t} \int_{-\infty}^{s} P( \Theta ,t \mid y) dy
\end{align*}
Based on this equation we can write a recursion formula for moments of  FPT with diffusion strength $B(y)$ and drift term $A(y)$:

\begin{align} \label{eq:app4-1}
t_n(\Theta \mid x_0) = n \int_{x_0}^{\Theta} \dfrac{2 dz}{B(z) W(z)} \int_{-\infty}^{z} W(k) t_{n-1} (\Theta \mid k)dk \tag{A4.1}
\end{align}
with $t_0 =1 $, and  $W(k)$ is the stationary probability distribution
\begin{align} \label{eq:A4.2}
W(k) = \dfrac{C}{B(x)} exp [ \int dx \dfrac{2A(x)}{B(x)}]  \tag{A4.2}
\end{align}
In particular, for the first moment we have 
\begin{align} \label{eq:A4.3}
t_1(\Theta \mid x_0) = \int_{x_0}^{\Theta} \dfrac{2 dz}{B(z) W(z)} \int_{-\infty}^{z} W(x) dx \tag{A4.3}
\end{align}
 For the case of Langevin equation in section \ref{sec3-sub1-1}, writing $x_0 = (V_{leak} - V_{th}) + ( V_{th} - \dfrac{a}{b}) = - \Theta + x_{th} $ and  using equations \ref{eq:A4.2} and \ref{eq:A4.3}, the average first passage time can be written as 
 \begin{align*}
t_1(x_{th} \mid x_0) =& \dfrac{\sqrt{\pi}}{b} \int_{ x_0 \sqrt{\dfrac{b}{\sigma^2}}}^{x_{th}\sqrt{\dfrac{b}{\sigma^2}}} e^{z^2} (1 + erf(z)) \\
 = & \dfrac{\sqrt{\pi}}{b} [ \int_{0} ^{x_{th}\sqrt{\dfrac{b}{\sigma^2}}} e^{z^2} (1 + erf(z)) \\&
-  \int_{0} ^{(- \Theta + x_{th})\sqrt{\dfrac{b}{\sigma^2}}} e^{z^2} (1 + erf(z)) ] \\
\end{align*}

where $erf(.)$ is the Gauss error function.

Taking $ y = x_{th} \sqrt{\dfrac{b}{\sigma^2}} $ and $\theta = \Theta
\sqrt{\dfrac{b}{\sigma^2}} $  and using the following series expansion for small $y$ 
\begin{align} \label{eq:App4}
 & \int_{0}^{y} e^{z^2} dz =  y + \dfrac{y^3}{3} + \dfrac{y^5}{10} ... \nonumber \\
 & \int_{0}^{y}  e^{z^2}erf(z) dz = \dfrac{1}{\sqrt{\pi}} ( y^2 + \dfrac{y^4}{3} + \dfrac{4 y^6}{45} ...) \nonumber  \\
 & \int_{0}^{-\theta + y}  e^{z^2}(1 +erf(z)) dz = \int_{0}^{-\theta}  e^{z^2}(1 +erf(z))dz  \nonumber  + y ( e^{-\theta^2}(1 +erf(-\theta)) \nonumber  \\ 
 &  \approx  \int_{0}^{-\theta}  e^{z^2}(1 +erf(z))dz \equiv \kappa  \tag{A4.4}
\end{align}
we arrive at equation \ref{eq:fpt} of section \ref{sec3-sub1-1}.

 It can be seen from equation \ref{eq:app3-1} that the factor $ \sqrt{\dfrac{b}{\sigma^2}} $ in large balanced rates of excitatory and inhibitory inputs asymptotically approaches a constant value. Therefore, we can approximate  $\kappa$ to be very weakly dependent on input rates and take it as a constant factor.
 Next, we want to investigate the variance of first passage time. From the recursion formula \ref{eq:A4.1} we have
\begin{align*}
t_2( x_{th} \mid x_0 ) = & \dfrac{2\pi}{b^2} \int_{ -\infty} ^{x_{th}\sqrt{\dfrac{b}{\sigma^2}}} e^{z^2} (1 + erf(z))^2  \int_{z}^{x_{th}\sqrt{\dfrac{b}{\sigma^2}}} dr e^{r^2} \Theta (r -x_0 \sqrt{\dfrac{b}{\sigma^2}}) 
\end{align*}
After some straightforward calculations we arrive at
\begin{align*}
t_2( x_{th} \mid x_0 ) = & \dfrac{2 \sqrt{\pi}}{b} t_1(x_{th} \mid x_0) [ \int_{0} ^{x_{th}\sqrt{\dfrac{b}{\sigma^2}}} e^{z^2} (1 + erf(z))]  \\
+ & \dfrac{2 \sqrt{\pi}}{b^2}ln(2) [ \int_{0} ^{x_{th}\sqrt{\dfrac{b}{\sigma^2}}} e^{z^2} -  \int_{0} ^{x_{0}\sqrt{\dfrac{b}{\sigma^2}}} e^{z^2}] \\ 
- & \dfrac{2 \sqrt{\pi}}{b^2} [ \phi(x_{th}\sqrt{\dfrac{b}{\sigma^2}}) - \phi (x_{0}\sqrt{\dfrac{b}{\sigma^2}}) ] \\
 - &  \dfrac{2 \sqrt{\pi}}{b^2} [ \psi (x_{th}\sqrt{\dfrac{b}{\sigma^2}}) - \psi (x_{0}\sqrt{\dfrac{b}{\sigma^2}}) ] \\
\end{align*}
where functions $ \phi$ and $\psi$ are multi-variable integrals containing powers of  $erf$ and $e^{x^2}$ in the integrand with  series expansion :  

\begin{align*}
& \phi(y) = \sum_{n=0}^{\infty} \dfrac{y^{2n+3}} {(n+1)!(2n+3)} \sum_{k=0}^{n} \dfrac{1}{2k+1} \\
& \psi(y) = \sum_{n=0}^{\infty} \dfrac{2^n y^{2n+4}} {(n+2)(2n+3)!!} \sum_{k=0}^{n}\dfrac{1}{k+1} \\
\end{align*}

Rewriting the first term in brackets in terms of first passage time and considering just terms linear in $x_{th}$
\begin{align}\label{eq:app4-4}
t_2( x_{th} \mid x_0 ) = & 2 t_1(x_{th} \mid x_0) [ t_1(x_{th} \mid x_0)  \nonumber \\ +& \dfrac{\sqrt{\pi}}{b} \int_{0} ^{(- \Theta + x_{th})\sqrt{\dfrac{b}{\sigma^2}}} e^{z^2} (1 + erf(z)) ]  \nonumber  \\
+ & \dfrac{2 \sqrt{\pi}}{b^2}ln(2) [ \int_{0} ^{x_{th}\sqrt{\dfrac{b}{\sigma^2}}} e^{z^2} -  \int_{0} ^{x_{0}\sqrt{\dfrac{b}{\sigma^2}}} e^{z^2}] \nonumber  \\ 
- & \dfrac{2 \sqrt{\pi}}{b^2} [ \phi(x_{th}\sqrt{\dfrac{b}{\sigma^2}}) - \phi (x_{0}\sqrt{\dfrac{b}{\sigma^2}}) ] \nonumber  \\
 - &  \dfrac{2 \sqrt{\pi}}{b^2} [ \psi (x_{th}\sqrt{\dfrac{b}{\sigma^2}}) - \psi (x_{0}\sqrt{\dfrac{b}{\sigma^2}}) ] \nonumber  \\
 \approx & 2 t_1(x_{th} \mid x_0)^2 + \dfrac{2 x_{th} \sqrt{\pi}}{b\sqrt{b} \sigma}ln(2) \nonumber  \\ &+ C(x_{th} =0 )  \tag{A4.5}
\end{align}
where the constant $C$, coming  from the integral, is negative.

\end{appendices}


\end{document}